\documentclass[12pt,preprint]{aastex}
\usepackage{apjfonts}
\def\msun{$\rm M_{\sun}$}

\def\Rc{$\rm R_{C}$}
\def\Ic{$\rm I_{C}$}
\def\degree{$^{\circ}$}
\def\SOri{${\sigma}$ Orionis\space}

\begin{document}
\shortauthors{Hernandez et al.}
\shorttitle{Spectroscopic census in the \SOri cluster}

\title{A spectroscopic census in young stellar regions: the \SOri cluster}

\author{Jes\'{u}s Hern\'andez\altaffilmark{1}, Nuria Calvet \altaffilmark{2}, Alice Perez \altaffilmark{1,3},
Cesar Brice\~{n}o\altaffilmark{4}, Lorenzo Olguin\altaffilmark{5}, Maria E. Contreras\altaffilmark{6}, 
Lee Hartmann \altaffilmark{2}, Lori Allen\altaffilmark{7}, Catherine Espaillat\altaffilmark{8}, 
Ram\'{i}rez Hernan\altaffilmark{1}}

\altaffiltext{1}{Centro de Investigaciones de Astronom\'{\i}a, Apdo. Postal 264, M\'{e}rida 5101-A, Venezuela.}
\altaffiltext{2}{Department of Astronomy, University of Michigan, 500 Church Street, Ann Arbor, MI 48109, US}
\altaffiltext{3}{Departamento de F\'{\i}sica, Universidad de Oriente, Venezuela}
\altaffiltext{4}{Cerro Tololo Interamerican Observatory, Casilla 603, La Serena, Chile}
\altaffiltext{5}{Depto. de Investigaci\'{o}n en F\'{\i}sica, Universidad de Sonora, Sonora, M\'{e}xico}
\altaffiltext{6}{Instituto de Astronom\'{\i}a, Universidad Nacional Aut\'{o}noma de M\'{e}xico, Ensenada, BC, M\'{e}xico}
\altaffiltext{7}{National Optical Astronomy Observatory, 950 North Cherry Avenue, Tucson, AZ 85719, USA }
\altaffiltext{8}{Department of Astronomy, Boston University, 725 Commonwealth Avenua, Boston, MA 02215, USA}
\email{hernandj@cida.ve}

\begin{abstract}
We present a spectroscopic survey of the stellar population of 
the \SOri cluster. We have obtained spectral types for 340 stars. 
Spectroscopic data for spectral typing come from several spectrographs 
with similar spectroscopic coverage and resolution. More than a half of stars 
of our sample are members confirmed by the presence of lithium in absorption, 
strong H$\alpha$ in emission or weak gravity-sensitive features. 
In addition, we have obtained high resolution (R$\sim$34000) spectra in the H$\alpha$ 
region for 169 stars in the region. Radial velocities were calculated 
from this data set. The radial velocity distribution for members of the cluster 
is in agreement with previous work. Analysis of the profile of the 
H$\alpha$ line and infrared observations reveals two binary systems or 
fast rotators that mimic the H$\alpha$ width expected in stars with accretion disks. 
On the other hand there are stars with optically thick disks and 
narrow H$\alpha$ profile not expected in stars with accretion disks. 
This contribution constitutes the largest homogeneous spectroscopic 
data set of the \SOri cluster
to date.

\end{abstract}

\keywords{infrared: stars: formation - stars: pre-main sequence - 
open cluster and associations: individual ( $\sigma$ Orionis cluster) 
protoplanetary systems: protoplanetary disk}

\section{Introduction}
\label{sec:int}
\defcitealias{hernandez07a}{H07a}

The  \SOri cluster located in the Orion OB1 association was first recognized 
by \citet{garrison67} as a group of  15 B-type stars around the massive and 
multiple system $\sigma$ Ori AB. Low-mass members of the 
\SOri cluster were reported by \citet{wolk96} and \citet{walter97}, who found over 
80 X-ray sources and spectroscopically identified more than 100 low mass, pre-main
sequence (PMS) stars. A copious amount of work on this 
cluster have revealed several hundred low mass stars and brown dwarfs that 
could belong to the \SOri cluster \citep[see ][]{walter08}. In the Mayrit catalog, 
\citet{caballero08a} presents a compilation of 241 stars and brown dwarfs which have 
known features of youth (\ion{Li}{1}$\lambda$6708 in absorption, strong emission 
of H$\alpha$, infrared excess, X-ray emission or  weak gravity-sensitive features).
More than one third of these stars have disks identified using Spitzer photometry 
\citep{hernandez07a, luhman08}. Recently, deep and wide near infrared surveys  
have increased substantially the number of substellar candidates of the cluster 
\citep[e.g.,][]{lodieu09,bejar11,pena12}. The \SOri cluster has a dense core
extended from the center to a radius of 20{\arcmin}, in which most members are located,
and a rarefied halo extended up to 30{\arcmin} \citep{caballero08b}. Other general
properties of the \SOri cluster are described by \citet{walter08}.

The \SOri cluster is an excellent natural laboratory to study the formation
and early evolution of stars and protoplanetary disks in the entire range of
stellar masses, from their massive members to the lowest mass objects, such 
as brown dwarfs and free-floating planets.  It has a evolutionary stage at 
which the beginnings of disk evolution become evident and thus 
a large diversity of protoplanetary disks is observed \citep[][ hereafter H07a]{hernandez07a}. 
The age normally used for the cluster is 2-4 Myr \citep[e.g.,][]{zapatero02,oliveira04,
franciosini06,sherry08,bejar11,caballero12,rigliaco12,pena12}. Recently, 
\citet{bell13} have derived ages for 13 young stellar groups 
(including the \SOri cluster) that are a factor of two 
higher than the ages typically adopted in the literature. Regardless 
of the actual value of the age, a comparison of empirical isochrones reveals 
that the \SOri cluster must be younger than the $\lambda$ Orionis cluster, 
the Orion OB1b association, the $\gamma$ Velorum cluster and the 25 Ori cluster \citep{hernandez08};
the age of these groups typically adopted in the literature covers
the range from 5 to 10 Myr, and consistently, they have smaller protoplanetary
disk frequencies than $\sigma$ Ori.

The $\sigma$ Orionis cluster is relatively near and the reddening toward 
the center of the cluster is low \citep[E(B-V)$\lesssim$0.1 mag; ][]{brown94,bejar99,bejar04,sherry08}.
The distance of 352$^{+166}_{-85}$ pc calculated by Hipparcos for the central system ($\sigma$ Ori AB)
agrees, within the uncertainties, with the Hipparcos distance calculated
for the overall population of the stellar subassociation OB1b
\citep[439$\pm$33;][]{brown98}, which is statistically more reliable.
This value agrees to the distance of 420$\pm$30 estimated by \citet{sherry08} 
from main sequence fitting corrected for the sub-solar metallicity expected 
in the Orion OB1 association \citep{cunha98,walter08}. 
On the other hand,  assuming that $\sigma$ Ori AB is a 
double system, \citet{caballero08c}
estimated a distance of 334$^{+25}_{-22}$ for this object; 
however, he also estimated a distance of $\sim$385 pc 
in the case that $\sigma$ Ori AB 
was a triple system. Few years later, \citet{simon11} report a third 
massive star component confirming the $\sigma$ Ori AB as a triple system 
(with masses of 19M$_\sun$, 15M$_\sun$ and 9M$_\sun$). 
Finally, recent  interferometric observations 
of the $\sigma$ Ori system during periastron
in the Center for High Angular Resolution Astronomy
(CHARA) with the Michigan Infra-Red Combiner (MIRC) 
yield a distance of 380.7$\pm$7.1 pc for $\sigma$ Ori AB 
(Gail Schaefer and John Monnier, private communication). This value is 
in agreement with the distance estimated by \citep{caballero08c} 
and within 2$\sigma$ from the distance reported by \citet{sherry08}.   
{ In this paper, we assume a distance of 385 pc reported by \citet{caballero08c}.}

Regardless of the numerous studies in the \SOri cluster, dozens 
of candidates remain without spectroscopic confirmation of 
membership \citep{caballero08}, and spectroscopically determined 
stellar characterization exists for only a small fraction of 
the \SOri objects \citep{rigliaco11,rigliaco12,cody10}. 
Moreover, spectral types or effective temperatures published 
for members or candidates of the \SOri cluster are obtained 
from spectroscopic data with different spectral coverage and 
resolution and applying different methods, which could 
introduce systematic differences between several samples  
\citep[e.g.,][]{wolk96, walter97, bejar99, zapatero02, barrado03, scholz04, gonzalez08, 
caballero06, caballero08,cody11,caballero12,rigliaco12}. 
The spectral type is key for deriving fundamental quantities such as visual extinction, 
temperature and luminosity, which in turn allow the determination of stellar radii, 
masses and ages.
In this contribution, we analyze several sets of spectroscopic data in order 
to obtain homogeneously spectral types and spectroscopic membership for stars in the \SOri cluster.
This paper is organized as follows. In Section \ref{sec:obs}, we describe the
observational data and the target selection in the cluster.
In Section \ref{s:res}, we discuss results from the low resolution spectroscopic 
analysis (\S\ref{sptclass}), the high resolution spectroscopic analysis (\S\ref{radialvelocity}),
memberships based on photometric data and proper motions analysis (\S\ref{s:admem}), 
and an overall membership analysis of stars studied in this work (\S\ref{sec:mem}).
In Section \ref{s:mips}, we revisit the disk population of the \SOri cluster  
and we compare our results with the disk census from \citetalias{hernandez07a}. 
In Section \ref{s:slow_accretor}, we discuss the relation between the infrared excesses 
of disk bearing stars and the accretion properties obtained from high resolution analysis 
of the H$\alpha$ line.  Finally, we present our conclusions in Section \ref{sec:conc}.

\section{Observations and Target Selection}
\label{sec:obs}

\subsection{Optical Photometry}
\label{sec:phot}
\subsubsection{OSMOS Photometry}
\label{osmos}
We obtained optical photometry (UBV\Rc\Ic) of the center of the cluster 
on December 24, 2011 using  the Ohio State Multi-Object Spectrograph 
(OSMOS) on the MDM 2.4 m  Hiltner telescope \citep{stoll2010, martini2011}.
We obtained two sets of images, one short exposure set (20, 15, 10, 5 and 5 seconds  
for U, B, V, {\Rc} and \Ic, respectively) and one long exposures set (3x200, 3x200, 3x150, 3x100 
and 3x100 seconds for U, B, V, {\Rc} and \Ic, respectively).
OSMOS has an all-refractive design that re-images a 20\arcmin diameter 
field-of-view onto the 4064x4064 MDM4K CCD with a plate scale of 0.273\arcsec pixel$^{-1}$.
The effective non-vignetted field of view is 18.{\arcmin}5 x 18.{\arcmin}5. 
We used 2 x 2 binning,  which gives a final plate  scale of 0.55\arcsec pixel$^{-1}$. 
The MDM4K CCD is read out using four amplifiers and has a known issue of 
crosstalk between the four CCD segments of each amplifier. When a CCD pixel 
is saturated, it creates spurious point sources in corresponding pixels 
on the other three segments. 

Each OSMOS frame was first corrected by overscan using the IDL program proc4k 
written by Jason Eastman for the Ohio State 4k CCD imager and modified to process 
OSMOS data. We then performed the basic reduction following the standard procedure 
using IRAF. We performed aperture photometry for the initial sample (\S \ref{sec:sample}) 
with the IRAF package APPHOT. The IRAF routine mkapfile was used to determine aperture
corrections for each filter. We used Landolt standard fields for photometric calibration 
in the Johnson-Cousin system \citep{landolt09,landolt92}. If a star was saturated in the 
long exposures images, then the short-exposure  measurements were used.
Non-detections in the combined long exposure images, saturated stars in short exposure images, and 
photometry affected by bleeding or crosstalk from saturated stars 
were removed. The saturation limit for the short exposure image 
is V$\sim$12 and the limit magnitude for the combined long exposure
image is V$\sim$23.

\subsubsection{Additional photometry}
\label{morephot}

Since OSMOS photometry only covers the central region of the cluster (the dense core), 
we completed the optical data set used in this paper 
using Johnson-Cousin photometry from different catalogs.

\noindent
a) The CIDA Variability Survey of Orion \citep[CVSO;][]{briceno05,mateu12}, which is being 
carried out since 1999 using the Jurgen Stock 1 m  Telescope with the QUEST I camera 
\citep[an array of 4x4 CCD detectors;][]{baltay02} 
in the Venezuelan National Observatory at Llano del Hato. The region studied 
in this work is covered almost completely 
in this multi-epoch survey. Only
a small region in the range of declination from -2.52\degree to -2.44\degree,
corresponding to the gap between CCDs in the QUEST I camera \citep{vivas04,hernandez07a},
lacks data. The saturation limit for this survey is V$\sim$13.5 
and the limit magnitude is V$\sim$19.5.

\noindent
b) \citet{sherry04} presented a BV\Rc{\Ic} survey of 0.89 square degrees around the \SOri cluster.
Observations were made with the 0.9 and 1.5 m telescopes at the Cerro Tololo Inter-American
Observatory (CTIO). The reported completeness limit is V=18 within 0.3{\degree} of the 
$\sigma$ Ori AB star and V=20 for regions more than 0.3{\degree} from the $\sigma$ Ori AB star. 
This catalog includes optical photometry for 234 likely members of the cluster. 

\noindent
c) The Cluster Collaboration's Photometric Catalogs collects photometric data  
for several young stellar associations and clusters. The catalogs were 
created using the optimal photometry algorithm described in \citet{naylor98} 
and \citet{naylor02}. For the \SOri cluster, \citet{kenyon05} and \citet{mayne07}
presented \Rc\Ic and BV\Ic photometry, respectively. These data sets were obtained 
using the Wide Field Camera on the 2.5 meters Issac Newton Telescope with the 
Harris BVR filters  and the Sloan i filter. The magnitudes were  calibrated 
using Landolt fields, thus  the final photometry is reported in the 
Johson-Cousin systems. Sources of the initial sample with optical photometry 
from these catalogs cover a brightness range from V$\sim$11.5 to V$\sim$23.5

\noindent
d) The All-sky Compiled Catalogue of 2.5 million stars \citep[ASCC-2.5 V3;][]{kharchenko09}.
This catalog collects B and V Johnson photometry mainly from 
Hipparcos-Tycho family catalogs.
The limiting magnitude is V$\sim$12-14, 
although the completeness limit (to 90\%) is V$\sim$10.5 mag. 
We augmented the V, B optical data set toward the brightest 
objects using the ASCC-2.5 V3 catalog.

{ Using stars in common between the different catalogs and OSMOS photometry, we find (after 3$\sigma$ clipping) 
that the V-band measurements are comparable within 0.5\%, 1.6\% and 1.7\% for the CVSO,  \citet{sherry08} and
the Cluster Collaboration Photometric Catalog, respectively. We do not have enough common stars between OSMOS
data and \citet{kharchenko09} catalog to do this comparison.}

\subsection{Initial sample}
\label{sec:sample}
The initial sample in this study includes all 2MASS sources \citep[4659 sources;][]{cutri03} in a region of
48\arcmin x48\arcmin centered at RA=84.7\degree and DEC=-2.6\degree. This region covers the 
field studied in \citetalias{hernandez07a} using the four channels of the InfraRed Array Camera 
\citep[IRAC;][]{fazio04}. The 2MASS catalog is complete down to J$<$15.8, which includes stars 
beyond the substellar limit expected for the \SOri cluster \citepalias[e.g., J$\sim$14.6;][]{hernandez07a}.
We compared sources from the 2MASS catalog and from the United Kingdom Infrared Telescope (UKIRT) 
Infrared Deep Sky Survey \citep[UKIDSS;][]{lawrance07,lawrance13}. 
When 2MASS sources have poor photometric quality 
flag ("U", "F" or "E") for the J magnitude, we use 
photometric measurements reported in UKIDSS catalog.   

Out of 31 UKIDSS sources with J$<$15.8 and without 2MASS counterparts, 
10 sources are galaxies based on the profile classification of UKIDSS and 16 sources are close visual binaries
not resolved in 2MASS images (2MASS reports photometry only for the brightest companion).    
From the remaining 5 sources, 4 stars are background candidates and only one star 
is a photometric member based on its location in the Z versus Z-J color magnitude diagram \citep{lodieu09}. 
This star has an optically thick disk \citepalias[SO 566;][]{hernandez07a} and the 
non-detection in 2MASS images suggests large variability (variability amplitude in J-band $\gtrsim$ 2.7). 
The star SO 566 was added to the initial sample of 4659 2MASS sources. Finally, there are 4 additional sources 
studied in \citetalias{hernandez07a} without 2MASS counterpart (SO 406, SO 336, SO 361 and SO 950). Since 
all these sources are fainter (J$>$17.9 in UKIDSS) than the 2MASS completeness limit, 
they were not included in this study.

\subsubsection{2MASS sources without optical photometry}
\label{no2mass}

Out of 4660 sources in the initial sample, 4444 sources (95.4\%) 
have optical counterparts (at least in filter V). 
Out of 216 sources without optical photometry, 
160 stars are fainter than the completeness 
limit of 2MASS catalog and were not included in this
study. Out of 56 stars brighter than the completeness 
limit of 2MASS catalog, 52 stars have photometric memberships reported by \citet{lodieu09}
and 4 stars are brighter than the Z-band limit used by \citet{lodieu09} 
for their membership criteria. They found 14 photometric candidates, 
35 background candidates, and 3 Galaxies based on the stellar profile 
classification  and color magnitude diagrams. 
Three of these photometric candidates are studied in \S\ref{sptclass},
the remaining 11 sources are detailed in the Appendix \ref{app1}.

\subsection{Compilation of Known Members}
\label{known members}

We compiled lists of spectroscopically confirmed members of 
the \SOri cluster on the basis that they exhibit \ion{Li}{1}$\lambda$6708
in absorption or they have radial velocities expected for 
the kinematic properties of the cluster (hereafter "spectroscopic known members").  
 
Using radial velocity  measurements, \citet{jeffries06} showed 
that young stars located in the general region of the cluster 
consist of two stellar groups kinematically separated by 7 km $s^{-1}$ 
in radial velocity and  with different mean ages and distances. 
One group has radial velocities from 27 km $s^{-1}$ to 
35 km $s^{-1}$, which is consistent with the radial velocity of 
the central star $\sigma$ Ori AB \citep[29.5 km $s^{-1}$; ][]{kharchenko07}. 
The other group has radial velocities from 20 km $s^{-1}$ to 27 km $s^{-1}$, and could be in front of 
the first group and could have a median age and distance similar to older stellar groups associated 
with the sparser Orion OB1a subassociation \citep[age$\sim$10 Myr; distance$\sim$326;][]{briceno07}.
This older and kinematically distinct stellar population (hereafter "the sparser stellar population") 
is located to the north-west of the central system  and more difficult to detect in radial 
velocity distributions of stars located near the center of the cluster \citep{sacco08}.  
Using the limit defined by \citet{jeffries06}, 
we included in our study 162 kinematic members of the \SOri cluster, with radial 
velocity between 27 and 35 km $s^{-1}$, from \citet{zapatero02}, \citet{kenyon05}, \citet{caballero06}, 
\citet{burningham05}, \citet{maxted08}, \citet{sacco08} and \citet{gonzalez08}. 

We also included as spectroscopic known members, 181 stars with \ion{Li}{1}$\lambda$6708 in absorption 
from \citet{wolk96}, \citet{alcala96}, \citet{zapatero02}, \citet{muzerolle03}, \citet{barrado03}, \citet{andrews04}, 
\citet{kenyon05}, \citet{caballero06,caballero06b}, \citet{gonzalez08}, \citet{sacco08}, and \citet{caballero12}. 
Since the presence of lithium is an indicator of youth in late type stars, our 
selection includes stars with reported spectral types K or M. Our list includes all late type stars 
with \ion{Li}{1}$\lambda$6708 in absorption compiled in the Mayrit Catalog \citep{caballero08a}. 

Additionally, we have compiled 168 X ray sources reported by
\citet{caballero08a}, \citet{skinner08}, \citet{lopez08}, \citet{caballero09}, \citet{caballero10a} or X ray
sources in the XMM-Newton Serendipitous Source Catalogue 2XMMi-DR3 (XMM-SSC, 2010) with 
source detection likelihood (srcML) larger than 8 (sources with scrML$<$8 may be spurious).
Since young stars are strong X-ray emitters, likely members of the \SOri clusters are X ray
stellar sources located above the ZAMS. 

Finally, another indicator of youth is the infrared excess present when stars are surrounded 
by circumstellar disks. Using Spitzer Space Telescope observations, we reported 114 photometric
candidates with infrared excesses \citepalias{hernandez07a}.

Youth indicators based on the presence of \ion{Li}{1}$\lambda$6708 in absorption, 
X ray observations and infrared excesses do not discriminate interlopers from the 
sparser stellar population. Regardless of this expected contamination, we define 
the sample of known likely members (hereafter "known members") compiling kinematic 
members based on radial velocities, stars with \ion{Li}{1}$\lambda$6708 in 
absorption, X ray sources above the ZAMS and stars with infrared excesses. 
Table \ref{t:knownmembers} shows information about the known member sample.

\subsection{Target Selection}
\label{s:photcan}

Using the photometric data compiled in \S \ref{osmos} and \S \ref{morephot} we
selected photometric candidates of the \SOri cluster. 
Although we have V, \Rc, \Ic, and J magnitudes for most stars in the initial sample, 
we use the V-J color because is the most complete for our sample and it has the greatest color 
leverage for constraining PMS populations. 

Figure  \ref{f:sel} shows the color-magnitude diagram V vs V-J
for sources in the initial sample with optical counterparts. 
We have estimated an empirical isochrone as the location of known 
members compiled in \S \ref{known members}; stars with infrared excesses,
young stars confirmed using the presence of \ion{Li}{1}$\lambda$6708 in absorption, 
kinematic members confirmed using radial velocities, 
and X ray sources 
above the ZAMS \citep{sf00} located at 385 pc \citep{caballero08c}.
The empirical isochrone 
in Figure \ref{f:sel}
was estimated using the median V-J color 
of the known members for 1 mag bins in the V band. Standard deviations ($\sigma$) 
were calculated using the differences between the observed colors and the expected 
colors from the empirical isochrone. We fitted two straight lines to the points 
representing by the empirical isochrone color + 3$\sigma$ and by the empirical 
isochrones color - 3$\sigma$.
Photometric candidates are 
stars that fall between these lines \citep{hernandez08,hernandez10}. 

Our photometric sample includes the photometric candidates reported 
in \citetalias{hernandez07a} using a more limited optical photometric data set. 
The present sample is larger because we have included an updated version 
of the CVSO catalog \citep{mateu12}, photometry from the cluster collaboration and the new 
OSMOS photometry described in \S \ref{osmos}.  

Out of 255 spectroscopic known members and disk bearing candidates, 
249 stars are located in the photometric candidates region. 
Based on this, { we lose about 2.4\% of known members 
of the \SOri cluster with our photometric selection. 
Since targets for spectroscopic follow up generally are selected
using photometric cuts, the exclusion percentage of know members
could be a lower limit of the percentage of actual members excluded 
from our photometric selection. On the other hand, \citet{burningham05}
found that photometric selection cuts do not miss significant numbers
of bona-fide members of the \SOri cluster and thus the percentage of actual 
member excluded by our photometric selection can not be much higher than $\sim$2.4\%.}
Since our photometric 
criteria to select candidates is conservative, contamination by 
background stars is present in the entire color range.
At a color V-J$\sim1.5$, where a branch of old field stars 
crosses the PMS of the \SOri cluster, the expected contamination level 
by non-members of the cluster is quite high \citep[e.g.,][]{hernandez08}.
Table \ref{t:photcan} includes optical magnitudes for 
the photometric candidates and for the known members 
not located in the photometric candidates region.

\subsection{Low-Resolution Spectroscopy}
\label{s:spec}

Optical low-resolution spectra is a fundamental tool to identify
and characterize young stars by
using indicators as the presence of lithium in 
absorption, strong H$\alpha$ in emission or 
weak gravity-sensitive features. We have obtained low-resolution 
spectra for 340 stars located in the the \SOri cluster. 
This spectroscopic data comes from several spectrographs with similar 
spectral coverage and resolution (see Table \ref{t:spectrograph}). 
Targets for spectroscopic follow-up were selected mainly from the 
sample of photometric members with infrared excesses or the sample of 
photometric members with X-ray counterparts. Except for observations
obtained using the Hectospec multifibers spectrograph, 
target selection includes stars with V magnitude brighter than 16.5. 
Using the 3 Myr isochrone from \citet{sf00} and assuming a distance 
of 385 pc \citep{caballero08c} without reddening, this limit corresponds 
to stars of spectral type M3 or earlier (M$_*$$>$0.35\msun).

\subsubsection{Hectospec}
\label{s:hsp}

We obtained low resolution spectra with the fiber-fed multiobject
Hectospec instrument mounted on the 6.5m Telescope of the MMT Observatory (MMTO).
Hectospec has 300 fibers that can be placed within a 1{\degree} diameter field. 
Each fiber subtends 1.5\arcsec on the sky \citep{fabricant05}. The observations were taken with 270 line mm$^{-1}$ grating, 
providing $\sim$ 6 {\AA} resolution and spectral coverage of 3650 - 9200 {\AA}. 
The Hectospec data were reduced through the standard Hectospec data reduction pipeline \citep{mink07}.
This pipeline assumes that the sky background does not vary significantly with position 
on the sky and uses combined sky fibers (fibers that point to empty portions of the sky) 
to correct for the sky background. In general, the \SOri cluster has a smooth 
background. However, there are some targets located in regions with variable sky background 
and therefore their spectra could have bad estimates of nebular lines. 
{ We have identified 16 stars with bad sky subtraction and for these stars we did not
report measurements of the H$\alpha$ line. }
One observation was obtained on the night of 2006 October 11, which 
provided spectra for 160 photometric candidates and 56 additional stars in the field.

\subsubsection{FAST}
\label{s:fast}

We obtained low-resolution spectra for 46 photometric candidates using the
1.5 m telescope of the 
Fred Lawrence Whipple Observatory (FWLO) with the FAST 
spectrograph \citep{fabricant98}, equipped with the Loral 512x2688 CCD. 
The spectrograph was set up in the standard configuration used for FAST COMBO 
projects, a 300 groove mm$^{-1}$ grating and a 3\arcsec wide slit. 
This combination offers 3400{\AA} of spectral coverage centered at 5500 \AA, 
with a resolution of 6 \AA. This data were observed as part of the program
"Orion PMS Candidates" \citep[\#112;][]{briceno05,briceno07}. The spectra were reduced at the Harvard-Smithsonian
Center for Astrophysics using software developed specifically
for FAST COMBO observations. Additionally, we have five stars collected as part of the
FAST program "Ae/Be stars in OB associations" \citep[\#89;][]{hernandez05}. 
Finally, using the FAST Public Archive \footnote{http://tdc-www.harvard.edu/cgi-bin/arc/fsearch}, 
we have collected spectra for 19 additional stars observed as part of different programs. 

\subsubsection{OSU-CCDS}
\label{s:ccds}
Low resolution, long slit-spectra were obtained for 36 photometric candidates using the 1.3 m 
McGraw-Hill Telescope of the MDM Observatory (MDMO) with the OSU (Ohio State University) 
Boller \& Chivens CCD spectrograph (OSU-CCDS) equipped with the Loral 1200x800 CCD. 
We used the 158 grooves per mm grating centered at 5300 {\AA} along with a 1'' slit width . 
This configuration provides an effective resolution of 6.5 {\AA} with 3400{\AA} of spectral 
coverage (the nominal resolution of 4.1{\AA} provides spectra under sampled at the 1.3-m telescope). 
Observation were obtained on the nights of 2011 December 16 and 17. 
Processing of the raw frames and calibration of the spectra were carried out
following standard procedures using the IRAF packages twodspec and onedspec. 

\subsubsection{Boller \& Chivens - SPM}
\label{s:spm}

Low resolution, long slit-spectra were obtained with the Boller \& Chivens spectrograph 
mounted on the 2.1m telescope at the San Pedro Martir Observatory (OAN-SPM) 
\footnote{The Observatorio Astron\'{o}mico Nacional at  San Pedro Martir (OAN-SPM) 
is operated by the Instituto de Astronom\'{\i}a of the Universidad Nacional Aut\'onoma de M\'exico.}
during three observing runs: 2011 Oct 28 to 30, 2012 Oct 18 to 21 and 2013 Jan 12 to 14. A Marconi CCD 
(13.5 $\mu$m~pix$^{-1}$) with 2k x 2k pixel array was used as detector. We have used a
400 lines mm$^{-1}$ dispersion grating along with a 2{\arcsec} slit width, giving a spectral 
resolution of $\sim$6 {\AA}. Spectra reduction was carried out following standard procedures in 
XVISTA \footnote{XVISTA was originally developed as Lick Observatory Vista. It is currently 
maintained by Jon Holtzman at New Mexico State University and is available at
http://ganymede.nmsu.edu/holtz/xvista.}. In order to eliminate cosmic rays from our spectra, 
we have obtained three spectra for each star and combined them to get a median individual 
spectrum. A total of 34 photometric candidates were observed with this instrument.

\subsubsection{Boller \& Chivens - Cananea}
\label{s:can}
A set of low resolution, long-slit spectra was obtained with the Boller \& Chivens spectrograph 
mounted on the 2.1m telescope at the Observatorio Astrofisico Guillermo Haro (Cananea, Mexico) 
\footnote{The Observatorio Astrof\'{\i}sico Guillermo Haro (OAGH) is operated by Instituto Nacional
de Astrof\'{\i}sica Optica y Electronica (INAOE).}, during the nights 1-4 December 2012. 
A SITe CCD with 1k x 1k pixels of 24 microns was used as a detector. We have used a 150 lines 
mm$^{-1}$ dispersion grating and a slit width of 2\arcsec, giving  a spectral resolution of 
$\sim$10 {\AA}. Observing strategy and spectra reduction were the same as described in 
\S \ref{s:spm}. A total of 27 photometric candidates were observed with this instrument.

\subsection{High resolution spectroscopy}
\label{s:hectochelle}

We obtained high resolution spectra of a subset of candidates of the \SOri cluster 
using the Hectochelle fiber-fed multiobject echelle spectrograph
mounted on the 6.5 m Telescope of the MMT Observatory (MMTO).
Hectochelle can record up to 240 high resolution spectra simultaneously 
in a 1{\degree} circular field. Each fiber subtends 
1.5{\arcsec} on the sky \citep{szentgyorgyi11}.  
We use the order-sorting filter OB26 which provides a 
resolution of R$\sim$34000 with 180{\AA} of spectral coverage
centered at 6625\AA. In spite of the fact that the order-sorting 
filter OB26 is not optimal for radial velocity measurements \citep{furesz08},
the spectral coverage of this filter allows us to analyze the H$\alpha$ profile 
and the \ion{Li}{1}$\lambda$6708 line to identify young stars, accretors and non-accretors. 

One Hectochelle field was obtained on the night of 2007 February 27 
which provides high resolution spectra for 134 photometric candidates
and 8 additional stars in the region studied in this work. The data were reduced using an automated 
IRAF pipeline developed by G. Furesz which utilizes the standard spectral reduction
procedures, using IRAF and the tasks available under the packages mscred 
and specred. A more detailed description of Hectochelle data reduction can be found
in \citet{aurora06}. 

\section{Spectral types and membership}
\label{s:res}
\subsection{Spectral analysis, youth features and Reddening Estimates}
\label{sptclass}

Spectral types were derived applying the SPTCLASS code on the sample 
of stars with low resolution spectra (\S \ref{s:spec}). SPTCLASS is 
a semi-automatic spectral analyzing program that uses
empirical relations between spectral type and equivalent widths to
classify and characterize stars based on selected features \footnote{http://www.cida.gob.ve/$\sim$hernandj/SPTclass/sptclass.html}.
For spectral typing, it has three schemes optimized for different mass ranges 
(K5 or later, from late F to early K and F5 or earlier), which use different sets of 
spectroscopic features \citep{hernandez04,aurora05,downes08}. 
The user has to manually choose the best scheme for each star based on 
the prominent features in the spectrum and the consistency of several spectral indices.
The equivalent width (hereafter EW) for each spectral feature is obtained 
by measuring the decrease in flux due to photospheric absorption line from 
the continuum that is expected when interpolating between two adjacent bands. 
The spectral features measured by this procedure are largely 
insensitive to reddening and the signal to noise ratio (S/N) of the spectra 
(as long as we have enough S/N to detect the spectral feature). Moreover, spectral types
obtained by SPTCLASS are largely independent of luminosity because most 
of the indices selected are not sensitive to the surface gravity 
of the star \citep{hernandez04}. However, SPTCLASS does not take 
into account the effect on the lines of the hot continuum emission produced 
by the accretion shocks. This continuum emission makes the photospheric absorption 
lines appear weaker and for K-type and M-type highly veiled stars the SPTCLASS outputs 
should be considered as the earliest spectral-type limits \citep{tina12}. 
SPTCLASS is unable to give spectral types for highly veiled 
stars earlier than K-type \citep[e.g. Continuum Stars in][]{hernandez04}. 
In Table \ref{t:lowres} we report spectral types for the sample observed in
\S \ref{s:spec}.  

In Figure \ref{f:comp} we compare spectral types derived in this work with those
published previously from selected works \citep{wolk96,houk99,zapatero02,caballero06,caballero12,rigliaco12}.
Additionally, our spectral types were compared with spectral types compiled in SIMBAD
\footnote{SIMBAD, Centre de Donne\'{e}s astronomiques de Strasbourg, at http://simbad.u-strasbg.fr/simbad/} database
from different authors \citep{cohen79,nesterov95,houk99,zapatero02,barrado03,muzerolle03, hernandez05,briceno05,oliveira06,caballero06b,caballero10a,
caballero12, gatti08,sherry08,skinner08, renson09, naze09, cody10, townsend10, rigliaco11,rigliaco12}.
Most spectroscopic studies of the \SOri cluster have derived spectral types for low mass stars 
and very low mass stars (spectral types K5 or later). Thus, in Figure \ref{f:comp} we have few points of comparison
in the solar type range (from F to early K).
In general, and with the exception of \citet{wolk96}, spectral types reported previously by other authors, 
agree within the uncertainties, with spectral types calculated using SPTCLASS.  
\citet{wolk96} used absorption line ratios of \ion{Ca}{1} lines ($\lambda$6122, $\lambda$6162 and $\lambda$6494)
and \ion{Fe}{1} lines ($\lambda$6103, $\lambda$6200 and $\lambda$6574) as indicators of spectral types.
The spectral types dependencies of these indicators are weak for stars with spectral types K-early or earlier.
Additionally, the spectroscopic data set of \citet{wolk96} have poor signal-to-noise ratios,
which severely affects the measurements \citep{zapatero02}. A subset of the stars studied by 
\citet{wolk96} were observed by \citet{zapatero02}. Their spectral types, based on the measurements 
of molecular bands like TiO and CaH, match our determinations. On the other hand,
\citet{caballero06} used indices based on the EW of some lines (\ion{Fe}{1}$\lambda$6400, \ion{Ca}{1}$\lambda$6439, 
\ion{Ca}{1}$\lambda$6450, \ion{Ca}{1}$\lambda$6462 and \ion{Ca}{1}$\lambda$6717) to estimate 
spectral types. The spectral types dependencies of these indices are very weak 
in the spectral type range from G0 to K0, where we observe larger discrepancies between our spectral types
and the spectral types reported by \citet{caballero06}.

EWs of \ion{Li}{1}$\lambda$6708 and H$\alpha$ lines were measured using the 
interactive mode of SPTCLASS in which two adjacent continuum points
must be manually selected  on the spectra to fit a Gaussian line to the feature
\footnote{ In general, the spectroscopic features in which we obtain EW show Gaussian profiles}. 
We find that the typical uncertainties of EWs measured 
in this way are $\sim$10\%. In Table \ref{t:lowres} we assign a flag to each 
star based on the presence of \ion{Li}{1}$\lambda$6708 in absorption: 
"2" means that the star has \ion{Li}{1}$\lambda$6708 in absorption; 
"1" means that there are doubts about the measurement of \ion{Li}{1}, generally, 
due to low S/N of the spectra; 
"0" means that \ion{Li}{1}$\lambda$6708 is not present in the spectra.

Based on the EW of H$\alpha$ and the criterion 
defined by \citet{barrado03} to distinguish between accretors and non-accretors
for stars with spectral type G5 or later, in Table \ref{t:lowres} we classify stars 
into different groups: accretors (Acr), non-accretors (nAcr), and stars in which 
uncertainties for EWs of H$\alpha$ and spectral types fall on 
the limit between accretors and non-accretors (Acr?). We also have identified 
stars with spectral types earlier than G5 with H$\alpha$ in emission (ET$_{em}$) 
and with H$\alpha$ in absorption (E$_{abs}$). Figure \ref{f:CW_TTS} shows the relation 
between the EW of H$\alpha$ and the spectral type. In order to plot our data in a logarithmic scale,
EWs of H$\alpha$  have been shifted by ten units  (the dotted line is the limit between absorption 
and emission of H$\alpha$). Solid line delimits the area for accretors and non-accretor stars 
based on the EWs of H$\alpha$ emission line \citep{barrado03}. Using different symbols we 
plot results obtained using different spectrographs (see Table \ref{t:spectrograph}). 
We also plot stars bearing protoplanetary disks \citepalias{hernandez07a}. In general, stars
above the Barrado's accretion cutoff were classified as stars bearing full disks (II) 
or Transitional Disks (TD) in \citetalias{hernandez07a}. There are some stars with optically 
thick disks below { the H$\alpha$ EW criterion} (SO 435, SO 467, SO 566, SO 598, SO 682, SO 823 and SO 967).
{ These stars would be similar to CVSO 224 that was classified as a weak-line T Tauri star 
based on the EW of H$\alpha$. However, using high resolution spectra, \citet{espaillat08} show that is 
a slowly accreting T Tauri star with accretion rate of 7x10$^{-11}$ M$_\odot$ yr$^{-1}$ 
estimated using the magnetospheric accretion models \citep{muzerolle01}.} 
On the other hand, there are diskless star
that were classified as accretor (SO 229, SO 1123 and SO 1368).
These diskless objects with H$\alpha$ in emission could have substantial 
contribution produced by strong chromospheric activity which is related 
to fast rotation. In section \S\ref{s:slow_accretor}, we study in more detail
slowly accreting stars and diskless stars that mimic the H$\alpha$ width 
expected in stars with accretion disks.

For very low mass stars (VLMS) observed with HECTOSPEC, we can 
use the line \ion{Na}{1}$\lambda8195$ as an additional indicator of youth
\citep{downes08}. Since VLMS in the \SOri cluster are still 
contracting, their surface gravities are lower than that expected
for a main sequence star with similar spectral type.
The line \ion{Na}{1}$\lambda8195$ is the strongest one of 
the sodium doublet (\ion{Na}{1}$\lambda\lambda8183,8195$)  which is sensitive to surface gravity and 
varies significantly between M-type field dwarf and PMS objects \citep{schlieder12}.
The EW of the line \ion{Na}{1}$\lambda8195$ was estimated fitting a Gaussian
function to the feature on the normalized spectrum. We use the continuum bands suggested
by \citet{schlieder12}  to normalize each spectrum. The uncertainties in EWs 
were calculated as in \citet{kenyon05} using the scale of HECTOSPEC ($\sim$1.2\AA/pixel)
and the S/N of the continuum bands. 
Figure \ref{f:sodium_lines} shows 
normalized spectra of three stars with spectral type M5 illustrating the procedure 
to measure the line \ion{Na}{1}$\lambda8195$. 
The sources 05400029+0142097 and 05373723+0140149 are 
CVSO's stars spectroscopically confirmed 
as a dwarf field star and as a $\sim$10 Myr old star, 
respectively (Briceno et al in prep). 
The star SO 460 has \ion{Li}{1}$\lambda$6708 in absorption and has been confirmed previously 
as member of the \SOri cluster using radial velocity \citep{zapatero02,sacco08,maxted08}.  
It is apparent that the M type field star exhibits stronger
absorption in the sodium doublet in comparison to the PMS stars. 

Figure \ref{f:sodium_mem}  shows the EW of \ion{Na}{1}$\lambda8195$  
versus spectral type for stars in the \SOri sample,
 illustrating the procedure of using the sodium line as an additional 
criteria supporting membership in the VLMS of $\sigma$ Ori. 
We display the median and the second and third quartiles of \ion{Na}{1}$\lambda8195$ measured 
for a sample of PMS stars (lower solid line) and  for a sample of M-type field dwarfs (upper solid line) 
in the Orion OB1a and OB1b associations (Briceno in prep.). 
These PMS stars, with ages ranging from $\sim$5 to $\sim$10 Myr, were confirmed by the presence 
of \ion{Li}{1}$\lambda$6708 in absorption.    In general, stars confirmed as 
members by the presence of \ion{Li}{1}$\lambda$6708 in the \SOri cluster (open squares) are below
the median population of PMS stars in the OB1a and OB1b associations. On the other hand, stars selected as 
non members of the \SOri cluster (open circles) follow the median line of M-type field dwarfs in 
the OB1a and OB1b associations. Figure \ref{f:sodium_mem} shows that the separation between M-type 
field stars and PMS stars is more clear for later spectral types, 
so that for stars with spectral type M3 
or later we use the EW of \ion{Na}{1}$\lambda8195$ 
as an additional criteria of youth;
those stars located below the median line of the PMS stars 
in the OB1a and OB1b associations were identified as young stars
(labeled with "Y" in column 10 of Table \ref{t:lowres}). Most of 
the $\sigma$ Ori VLMS with uncertain membership based on \ion{Li}{1}$\lambda$6708 
(crosses) have smaller values of EW of \ion{Na}{1}$\lambda8195$ and thus are likely members 
of the cluster. The stars SO 576 and SO 795 exhibit strong \ion{Na}{1}$\lambda8195$ line 
and thus are classified as M-type field stars (labeled with "N" in Table \ref{t:lowres})

In order to estimate reddening for our sample, we calculate the root mean square
of the differences between the observed colors and the standard colors
 
\begin{equation}
RMS(A_V)=\sqrt{\frac{\sum{([V-M_\lambda]_{obs}-[1-\frac{A_\lambda}{A_V}]*A_V - [V-M_\lambda]_{std})^2}}{n}}
\end{equation}

where $[V-M_\lambda]_{obs}$ are the observed colors V-Rc, V-Ic and V-J (when they are available),
n is the number of colors used to calculate RMS($A_V$) and $[V-M_\lambda]_{std}$ are the corresponding 
standard colors for a given spectral type. 
The values of $\frac{A_\lambda}{A_V}$ were obtained from \citet{ccm89} assuming the extinction law normally
used for interstellar dust ($\frac{A_\lambda}{A_V}$=  0.832, 0.616 and 0.288 for Rc, Ic and J, respectively).
We vary the visual extinction ($A_V$)  from 0 to 10 magnitudes in steps of 0.01 magnitudes
and estimate visual extinction as the $A_V$ value with the lowest RMS$(A_V)$.
We use the intrinsic colors for main sequence stars from \citet{kh95} and the intrinsic 
colors of 5-30 Myr old PMS stars from \citet{pecaut13}. Since these PMS intrinsic colors 
are given for stars F0 or later, we complete the standard colors for stars earlier than F0 using 
the intrinsic color of 09-M9 dwarf stars compiled by \citet{pecaut13}. 
Intrinsic colors of PMS stars earlier than G5 appear to be consistent 
with the dwarf sequence \citep{pecaut13}.

Since PMS stars are in a different evolutionary stage than dwarf field stars, 
there may be systematic errors in the calculation of interstellar reddening 
when using main sequence calibrators. Figure \ref{f:avcomp} shows a comparison 
between the reddening calculated using 
the intrinsic colors for main sequence stars from \citet{kh95}  and the intrinsic colors 
for PMS stars from \citet{pecaut13}. 
In general, for stars in the spectral type range from G2 to M3 
the visual extinctions estimated from \citet{kh95} 
are larger than those estimated using PMS colors \citep{pecaut13}. On the other hand, for stars 
later than M3 (open squares), visual extinctions estimated from \citet{kh95} are slightly smaller 
than those estimated using PMS colors \citep{pecaut13}. 

Sorted by spectral types, Table \ref{t:lowres} summarizes  the results obtained 
from the low resolution dataset. Column (3) indicates the spectrograph used 
for each star (see Table \ref{t:spectrograph}). Column (4) shows spectral types. 
Columns (5), (7) and (9) show the EWs of \ion{Li}{1}$\lambda$6708, H$_\alpha$
and \ion{Na}{1}$\lambda8195$, respectively. Flags based on these lines are shown in columns (6), 
(8) and (10). Columns (11) and (12) show visual extinctions calculated using standard colors for
main sequence stars \citep{kh95} and using standard colors for PMS stars \citep{pecaut13}.

\subsection{Radial Velocity, H$_\alpha$ and \ion{Li}{1}$\lambda$6708 measurements}
\label{radialvelocity}

We measured heliocentric radial velocities (RVs) in the high resolution spectra of the sample
observed with Hectochelle. RVs were derived using the IRAF package "rvsao" which cross-correlate each 
observed spectrum with a set of templates of known radial velocities \citep{tonry79,mink98}.
We used the synthetic stellar templates from \citet{coelho05} with solar metallicity, 
surface gravity of log(g) = 3.5 and effective temperature ranging from 3500 to 7000 K in steps of 250 K.
The use of synthetic templates enables us to explore a wider range of stellar parameters than 
a few observed templates \citep{tobin09}. We estimated the radial velocity and the template as those 
that gave the highest value of the parameter R, defined as the S/N
of the cross correlation between the observed spectrum and the template. 
The RV error is calculated from the parameter R as \citep{tobin09}: 

\begin{equation}
 RV_{err}=\frac{14 \space (km s^{-1})}{1+R}
\end{equation}

In Table \ref{t:hectochelle} we report radial velocities for 95 stars in which R is larger 
than 4 (Rverr$\lesssim$2.8 km $s^{-1}$). This sample includes 36 stars with known radial velocities 
compiled in \S \ref{known members}. In general, the differences between our measurements and known radial velocities were 
less than 2 km $s^{-1}$ (rms=1.9 km $s^{-1}$). There are 5 stars (SO 616, SO 662, SO 791, SO 929 and SO 1094) 
with differences larger than 2 km $s^{-1}$. Since during its orbit around the center of mass 
radial velocities of the components of binaries could change, these 5 stars are labeled as binary candidates 
by radial velocity variability (RVvar). The cross correlation function also can be used to identify 
double lined spectroscopic binaries (SB2). We have identified 4 SB2 stars that show
double peaks in the cross correlation function.

{ EWs of the \ion{Li}{1}$\lambda$6708 line reported in Table \ref{t:hectochelle} 
were calculated fitting a Gaussian function to the line in the high resolution spectra. 
The errors of EWs were calculated as in \citet{kenyon05} using the scale of HECTOCHELLE (0.0786"/pixel) 
and  the S/N of the continuum used to obtain the EW. 
Figure \ref{f:RvLi} shows a diagram of EWs of \ion{Li}{1}$\lambda$6708 versus radial velocities.}
The distribution of radial velocities can be described by a Gaussian function centered
at 30.8 km $s^{-1}$ with a standard deviation ($\sigma_{RV}$) of 1.7 km $s^{-1}$. Applying a 3 $\sigma_{RV}$ criteria, we can define 
a radial velocity range for kinematic members of the cluster from 25.7 km $s^{-1}$ to 35.9 km $s^{-1}$. These values are similar to those 
reported by \citet{jeffries06} for the stellar group associated with the star $\sigma$ OriAB (27 - 35 km $s^{-1}$). 
In Table \ref{t:hectochelle} we label as kinematic members of the \SOri cluster stars with radial velocity 
in the membership range (flag "2") and as kinematic candidates of the sparser stellar population
(see \S \ref{known members}) stars with radial velocities from 20 km $s^{-1}$ to 25.7km $s^{-1}$ (flag "1").  
Stars with radial velocity smaller than 20 km $s^{-1}$ and larger than 35.9 km $s^{-1}$
were flagged as kinematic non-member (flag "0"). Three stars with \ion{Li}{1}$\lambda$6708 in absorption fall in the 
last group. The stars SO 362 and SO 1251 have RV>35.9 km $s^{-1}$ and the star SO 243 has RV<20 km $s^{-1}$.
The star SO 362 has a protoplanetary disk and its discrepant RV value could be caused by binarity.
The \ion{Li}{1}$\lambda$6708 measured in the star SO 1251 is lower than those expected for the \SOri stellar 
population and thus its membership is uncertain. Since the spectral type of the star SO 243 
is relatively early ( G5; see \S \ref{sptclass}), the youth of this object is also uncertain.
The shallow depth of the convective zone in stars earlier than K0 can allow them to reach the main 
sequence with a non-negligible amount of their primordial lithium content and thus \ion{Li}{1}$\lambda$6708 in absorption 
is not a reliable indicator of the PMS nature of these stars. Therefore, the presence of lithium in 
absorption in F type  and G type stars is only evidence that these objects are not old disk, post main 
sequence stars \citep{briceno97}. Finally, we were unable to calculate radial velocities for 17 stars 
with \ion{Li}{1}$\lambda$6708 in absorption. These stars were included as spectroscopic members of the cluster.

The full width of  H$_\alpha$ at 10\% of the line peak (WH$_\alpha$\_10\%)
is an indicator that allows to distinguish between accreting and non accreting
young stellar objects. \citet{white03} suggested that WH$_\alpha$\_10\% $>$270 km $s^{-1}$ 
indicates accretion independent of the spectral type. \citet{jayawardhana03} adopted a 
less conservative accretion cutoff of 200 km $s^{-1}$ studying young very low mass stars and 
brown dwarfs. A newly born star is surrounded by a primordial optically thick disk
which evolves due to viscous processes by which the disk is accreting material 
onto the star while expanding to conserve angular momentum \citep{hartmann98}.
As a consequence of this and other evolutionary processes, the frequency of accretors 
and the accretion rate in the inner disks steadily decreases from 1 to 10 Myr 
\citep[e.g.][]{calvet05,williams11}.
Thus signatures of accretion can be used as indicators of membership in a young stellar 
region. 
In the case in which H$\alpha$ has a symmetric or quasi-symmetric profile in emission, 
{ we measured automatically the peak of the profile fitting a Gaussian function to the H$\alpha$ 
line. When the fitting of a single Gaussian function does not work (e.g. non-symmetric profiles), 
the peak of the H$\alpha$ line was selected manually. The WH$_\alpha$\_10\%  was measured as 
the width of the H$\alpha$ velocity profiles at 10\% of the emission peak level (column 6 in Table 5).}
Following the criterion from 
\citet{white03}, stars with WH$_\alpha$\_10\% $>$270 km $s^{-1}$ were identified as accretors.
Stars without measurements of WH$_\alpha$\_10\% have comments in Table \ref{t:hectochelle}
about the H$\alpha$ profile.

\subsection{Additional criteria for membership}
\label{s:admem}
Since the criteria used in \S \ref{sptclass} as 
indicators of youth are useful mainly for stars with spectral types K and M, we need to use additional
criteria to confirm or reject candidates as member of the \SOri cluster. In general, 
the memberships described in this section are less reliable than those obtained from
low resolution spectroscopic analysis (\S \ref{sptclass}) and those 
obtained from high resolution spectroscopic analysis (\S \ref{radialvelocity})  .

\subsubsection{Photometric membership probabilities and variability}
\label{s:phot_prob}

Pre-main sequence (PMS) stars are characterized 
by having a high degree of photometric variability of diverse
nature \citep{herbst94,cody10,cody11}. The variability is due to chromospheric and magnetic
activity in the stellar surface (e.g., cool spots, flares and coronal mass ejections) 
and to protoplanetary disks around the stars (e.g., hot spots produced by accretion shocks,
variable accretion rates and variable extinction produced by inhomogeneities in the 
dusty disk). Since the expected photometric variability in PMS stars are 
so diverse, it is difficult to apply variability criteria such as range of periods or 
light curves types to refine the selection of possible members of a young stellar group. 
However, we can use the location of variable stars in color magnitude diagrams to
select them as PMS candidates.  \citet{briceno01,briceno05} indicate that selecting variable 
stars above the ZAMS located at the typical distance of a young stellar group 
clearly picks a significant fraction of members of that stellar group. Moreover, 
using the differences between the observed colors and the expected colors defined 
by empirical or theoretical isochrones, we can calculate photometric membership 
probabilities for the sample.

Figure \ref{f:CMD_rot} shows the procedure to estimate photometric membership probabilities. Based on [V-J] colors 
and V magnitudes, we tailored new indices (C$_{rot}$ and M$_{rot}$) using the following equations to rotate the color
magnitude diagram of Figure \ref{f:sel}:

\begin{equation}
 C_{rot}=[V-J]*cos(\theta)-(V-V_0)*sin(\theta)
\label{Rot1}
\end{equation}

\begin{equation}
 M_{rot}=V_0+[V-J]*sin(\theta)+(V-V_0)*cos(\theta)
\label{Rot2}
\end{equation}
where V$_0$ is the V value where the empirical isochrone has [V-J]=0
and $\theta$ is the angle where the root mean square of C$_{rot}$ for the known member sample is minimal ($\sim$25.5\degree). 
The distribution of C$_{rot}$ for known members (open circles) describes a Gaussian function with 
standard deviation of $\sigma$=0.32. Assuming a standard normal distribution we can transform C$_{rot}$ values to 
probabilities for our sample. The 3 $\sigma$ criterion applied in \S \ref{s:can} to select photometric 
candidates means that stars that fall outside of this criteria have membership probabilities lower 
than $\sim$0.3\% (dotted lines in Figure \ref{f:CMD_rot}). 
{ Notice that the photometric probability obtained here does not take into 
account the distribution of non-members in Figure \ref{f:CMD_rot}. 
The contamination by non-members is much higher at negative values of 
C$_{rot}$ than at positive values of C$_{rot}$. Also, the expected contamination is higher at
M$_{rot}\sim$13 where a branch of old field stars crosses the stellar population of the \SOri cluster}.

We plot in Figure \ref{f:CMD_rot} variable stars reported in the CVSO 
catalog \citep{briceno05,mateu12}, stars flagged as variable in the cluster collaboration's photometric catalogs \citep{kenyon05,mayne07}, 
variable low-mass stars and brown dwarfs of the \SOri cluster \citep{cody10},
and stars listed as known or suspected variables
 in The AAVSO International Variable Star Index \citep{watson06}. 
In spite of the fact that this compilation of variable stars in the \SOri cluster is not complete, 60\% of the stars
within the one  $\sigma$ criteria (dotted lines; photometric probability higher than $\sim$32\%) 
are reported as variable objects.

\subsubsection{Proper motion and { spatial distribution}}
\label{s:propermotion}

In general, the motion of young stellar groups of the Orion OB1 association 
are mostly directed radially away from the Sun. Thus, the expected intrinsic 
proper motions in right ascension ($cos(\delta)*\mu_\alpha$) and declination ($\mu_\delta$)
are small and comparable to measurement errors \citep{brown98,hernandez05}. 
Particularly, \citet{brown98} reported an average proper motion for this stellar 
association of 0.44 mas/yr and -0.65 mas/yr for $cos(\delta)*\mu_\alpha$ and $\mu_\delta$, 
respectively. Although we cannot use proper motion to separate potential cluster members 
from field star non-members, we can use criteria based on proper motions
to identify and reject high proper motion sources as potential members of the \SOri 
cluster \citep[e.g.,][]{lodieu09,caballero10}. 

We follow a similar method used in \citet{hernandez09} to identify potential 
non-members of the cluster. Figure \ref{f:PM}  shows the vector point 
diagram for the photometric candidates with proper 
motions reported in The fourth US Naval Observatory CCD Astrograph Catalog \citep[UCAC4;][]{zacharias13}.
The distributions of proper motions for the known members (green X's) can 
be represented by Gaussians centered at $cos(\delta)*\mu_\alpha \sim$ 2.2 mas/yr 
and $\mu_\delta \sim$0.7 mas/yr, with standard deviations of 
5.3 mas/yr and 4.1 mas/yr, respectively. These values agree within the errors 
with previous estimations for the entire OB association \citep[e.g.,][]{brown98}
and for the cluster \citep[e.g.,][]{kharchenko05,caballero07,caballero10}.
Most known members ($\sim$86\%) are located in a well defined region represented for the 3$\sigma$ 
criteria (dotted ellipse) in the vector point plot. We also use a less conservative criteria
of 5$\sigma$ (solid line ellipse) similar to the criteria used in previous works \citep{caballero07,caballero10,lodieu09}.
With the exception of the stars SO 592, SO 936 and SO 1368, the 5$\sigma$ criteria includes all the known member sample. 
The star SO 592 is a radial velocity member with \ion{Li}{1}$\lambda$6708 in absorption (see \S \ref{s:hectochelle}).
The star SO 1368, reported as diskless accretor in \S \ref{sptclass}, has large error in $cos(\delta)*\mu_\alpha \sim$ 
and thus the proper motion criteria for this object is uncertain.  
The star SO936 lies outside the vector point diagram. Its infrared excess at {8\micron} \citepalias{hernandez07a} 
is the only youth feature reported for this photometric candidate. 
Astrometric follow-up could help to find out if those stars are members of binary systems,
they were ejected from the cluster early on during the formation process, or they
belong to a moving group associated with Orion \citep{lodieu09}.

Based on the distribution of stars in the vector point diagram, we have classified our 
photometric candidates into three groups, stars inside of the 3$\sigma$ limit 
(proper motion flag "2"), stars between the 3$\sigma$ and the 5$\sigma$ limit 
(proper motion flag "1") and stars with proper motions larger than the 5$\sigma$ 
criteria (proper motion flag "0"). In general, the proper motion flags agree 
with previous proper motion studies to identify high proper motion interlopers 
\citep{lodieu09,caballero10}. Out of 29 stars studied by \citet{caballero10}
and located in the region studied in this work, 14 stars have proper motion 
in UCAC4. All the 13 photometric candidates identified by 
\citet{caballero10} as high proper motion interlopers have proper motion flag "0".
The other star is a background star (\S \ref{s:photcan}) rejected 
using optical colors \citep[star \#38 in ][]{caballero10} . Only one 
star (05401975-0229558) reported by \citet{lodieu09} as proper motion 
non-member have proper motion flag "1".  Additional studies are necessary  
to obtain youth features of this object.

{ Finally, \citet{caballero08b} suggests that the \SOri cluster has two components:
a dense core that extends from the center to a radius of 20\arcmin and a rarefied
halo at larger separations. Members of the \SOri cluster have higher probability
to be located in the dense core than in the rarefied halo. Thus, we also 
include a flag for the distance from the center of  the cluster. Stars located 
closer than 20\arcmin have flag "1", otherwise have flag "0".}  

\subsection{Membership analysis of the cluster}
\label{sec:mem}

Depending on the spectral type range, obtaining memberships of stars that belong to
a young stellar group can be a difficult process and some times we need to combine different
membership indicators. Based on the spectroscopic and photometric analysis developed in 
\S\ref{s:photcan}, \S\ref{s:mips}, \S\ref{radialvelocity}, \S\ref{sptclass} and \S\ref{s:admem}, and the 
information compiled in \S\ref{known members} about several membership criteria, we 
compile in Table \ref{t:mem1} membership indicators for stars with spectral types obtained in \S\ref{sptclass}. 
Columns (1) and (2) show source designations from \citetalias{hernandez07a} 
and \citet{cutri03}, respectively. Column (3) shows the spectral type. Column (4) indicates the 
membership flag based on the presence of \ion{Li}{1}$\lambda$6708 in absorption from low resolution
spectra (\S\ref{sptclass}) and from high resolution spectra (\S\ref{radialvelocity}). Columns (5)
shows the references of known members based on \ion{Li}{1}$\lambda$6708. Columns (6) and (7) show 
the radial velocity information from our analysis (\S\ref{radialvelocity}) and from previous works, respectively. 
Column (8) shows the classification based on the H$\alpha$ line and the accretion criteria 
from \citet[][\S\ref{sptclass}]{barrado03} and from \citet[][\S\ref{radialvelocity}]{white03}.
Column (9) indicates the membership flag based on the line \ion{Na}{1}$\lambda8195$ (\S\ref{sptclass}).
Additional membership information based on the presence of protoplanetary disks, X ray emission,
proper motion, distance from the center of the cluster and variability are in columns (10), (11), (12),
(13) and (14), respectively. Column (15) gives the photometric membership probability calculated 
in \S\ref{s:admem} and column (16) indicates the visual extinction estimated using the PMS standard colors
of \citet[][\S\ref{sptclass}]{pecaut13}. Finally, our membership flags and comments are given 
in the last column. Similar to the membership study by \citet{kenyon05}, some of our membership flags 
are arguable, but the reader can reach their own conclusions based on the information in Table \ref{t:mem1}.

In general, stars with spectral types B or A in Table \ref{t:mem1} were included 
as young stars by \citet{caballero07} and \citet{sherry08} using proper motion, radial velocity,  
X ray and infrared observations and main sequence fitting analysis. The star SO 602 was not included in those
studies. This star has discrepant radial velocity and very high visual extinction to be considered 
member of the \SOri cluster. The star SO 147 was not included in \citet{sherry08} and was rejected
by \citet{caballero07} based on the proper motions reported by \citet{perryman97}. SO 147 is a 
X ray source  and our RV measurements indicate that could be a young star member of the sparser population 
\citep{maxted08} or a binary of the \SOri cluster. However, the visual extinction of SO 147 is higher 
than that expected for the cluster, thus its membership is uncertain. \citet{sherry08} suggested that 
SO 956 is too bright to be member of the cluster and SO 521 is too faint to be a member 
of the cluster. The star SO 956 has youth features like infrared excess and X ray emission and the star 
SO 521 is located above the ZAMS but with relatively low photometric membership probability. 
Thus is not clear the membership status of these two objects. Finally, the star 
SO 139 is relatively bright to be a member of the cluster and \citet{sherry08} suggested that 
it could be a member of the cluster if this star is an equal mass binary.
In summary, the memberships of the stars SO 139, SO 956, SO 521 and SO 147 reported in Table 
\ref{t:mem1} are uncertain.

It is more difficult to estimate membership for solar type stars (F and G). 
In this spectral type range, old stars cross the sequence defined by the \SOri cluster 
and thus the contamination level by non-members of the cluster is quite high. 
For some stars \ion{Li}{1}$\lambda$6708 appear in absorption. However, for solar type stars 
this is not a robust criteria of youth and it is only evidence that these objects 
are not old disk, post main sequence stars \citep{briceno97,hernandez04}. Thus, we need 
to evaluate other membership indicators like presence of disks, X ray emission, 
variability, radial velocity proper motions, reddening and photometric membership 
probabilities to confirm or reject stars as members of the cluster. 
In Table \ref{t:mem1}, solar type stars with \ion{Li}{1}$\lambda$6708 in absorption 
have additional youth features that support their membership of the cluster 
(mainly X ray source, infrared excesses and variability). Four stars 
with \ion{Li}{1}$\lambda$6708 in absorption (SO 1123, SO 243, SO 92 and SO 379) have
discrepant radial velocity that makes us to suspect that they are not a member 
of the cluster. Another possibility is that these stars are binaries 
with variable radial velocities. Since, the frequency of binaries increases 
with the stellar mass \citep[e.g.,]{duchene13}, the probability to find a 
binary member with discrepant radial velocity in solar type stars is higher 
in comparison with low mass stars. We labeled these stars as probable members.
We rejected as members of the cluster stars with very low ($<$0.3\%) 
photometric membership probability and stars with discrepant radial velocity 
that do not have additional features of youth.  Finally, some stars do not 
have enough information to define their membership and were labeled as uncertain
members.

For low mass stars (from K0 to M2.5) and for very low mass stars (M3 or later)  the presence 
of \ion{Li}{1}$\lambda$6708 in absorption is a reliable indicator of youth. For very low mass stars
when the presence of \ion{Li}{1}$\lambda$6708 in absorption is uncertain (flag "1"), we used 
the measurements of \ion{NaI}{1}$\lambda8195$ to support the membership status. Moreover, 
stars with uncertain flag of \ion{Li}{1}$\lambda$6708 that exhibit other indicators to be member of the cluster
(including previous  \ion{Li}{1}$\lambda$6708 reported for the star) are also classified as members of 
the \SOri cluster. We also considered member of the cluster stars with protoplanetary disks
classified as accretors using the H$\alpha$ line. Finally, radial velocity can be used as a 
strong membership criteria for stars in the entire spectral type range and could be use to separate
the $\sigma$ Ori stellar population and the sparser stellar population (see \S \ref{known members}).

Optical spectra and SEDs for the entire sample with spectral types obtained 
in \S \ref{sptclass}, plus additional information
compiled from Tables \ref{t:photcan}, \ref{t:lowres} and \ref{t:mem1}
for each star, is available on-line \footnote{http://sigmaori.cida.gob.ve/}.
This online tool is a work in progress and we expect to add new objects 
and to extend the information about the population of the \SOri cluster.

Finally, Table \ref{t:mem2} compiles membership information for a set of stars studied 
in high resolution (\S\ref{radialvelocity}) but without low resolution analysis 
or spectral types (\S\ref{sptclass}). Our memberships are based on radial velocity
criteria, presence of \ion{Li}{1}$\lambda$6708 in absorption and the accretion 
criteria from \citet{white03}.

\section{Disk population and accretion}
\label{s:disk_accretion}

\subsection{Revisiting the disk census in the cluster}
\label{s:mips}

In \citetalias{hernandez07a}, we studied the disk population in 
the region covered by the four channels of IRAC.
Since mosaics of channels 4.5{\micron} and 8.0{\micron} have
a 6\arcmin.5 displacement to the north from the mosaics of 
channels 3.6{\micron} and 5.8{\micron}, the region with the complete 
IRAC data set is smaller than the field of view of individual 
IRAC mosaics. MIPS observations in the \SOri cluster also 
cover more area in the north-south direction in comparison 
to the region covered by the complete IRAC data set. A more 
detailed description of MIPS and IRAC data can be found
in \citetalias{hernandez07a}.

We searched for infrared excesses at 24{\micron}
in the newly identified candidates within the photometric sample (Table \ref{t:photcan})
reanalyzing the MIPS observation of the 
\SOri cluster. We followed the procedure described in 
\citet{hernandez07a,hernandez07b, hernandez08,hernandez09,hernandez10}
to identify stars with excess at 24\micron. Figure \ref{f:mips} shows 
the color-color diagram used to select new disk bearing candidates
of the cluster. 
Photospheric limits (dotted lines) were defined by \citetalias{hernandez07a} using 
the typical K-[24] color of disk less stars detected in the MIPS observation. 
An arbitrary limit between optically thick disks (II) and debris disks (dashed line)
was defined using the K-[24] color of a sample of known debris disks located in 
several young stellar groups \citep{hernandez11}.
Open squares represent stars not included in the previous MIPS 
analysis \citepalias{hernandez07a}. Out of 14  new disk bearing candidates, 12 have 24{\micron} 
infrared excess expected in stars with optically thick disks.  
The remaining two stars show 24{\micron} expected in stars with debris or evolved disks 
\citep[K-24$\lesssim$3.0; ][]{hernandez11}.
Two disk bearing candidates are potential galaxies based on the profile 
classification of UKIDSS (red X's). Spitzer photometry and disk type for each source 
are provided in Table \ref{t:disk_mips}. 
 
\subsection{Disk emission and Accretion Indicators}
\label{s:slow_accretor}

Independently of the spectral type, WH$_\alpha$\_10\%  is an indicator of accretion
\citep{white03,jayawardhana03}. Studying very low mass young objects, \citet{natta04} 
found that WH$_\alpha$\_10\% can be used to roughly estimate accretion rates (\.M$_{acc}$). 
The relation of \.M$_{acc}$ as a function of WH$_\alpha$\_10\% from \citet{natta04} indicates 
that stars accreting below the detectable level have log (\.M$_{acc}$) $\lesssim$ -10.3 M$_\odot$ yr$^{-1}$ 
and log (\.M$_{acc}$) $\lesssim$ -11 M$_\odot$ yr$^{-1}$ for the accretion cutoff of 270 km $s^{-1}$ 
\citep{white03} and 200 km $s^{-1}$ \citep{jayawardhana03}, respectively, although a large chromospheric 
contamination is expected if the mass accretion rates have such low levels \citep{ingleby11}. 

Figure \ref{f:EW10alfa} shows the relation between the WH$_\alpha$\_10\% versus 
the IRAC SED slope determined from the [3.6]-[8.0] color.  The horizontal solid 
line indicates the limit between optically thick disks and evolved disk
objects based on their infrared excess at 8{\micron} \citep{lada06, hernandez07a}.
 We plotted stars with optically thick disks (open circles), 
evolved disks (open squares) and transitional disks (open triangles) 
from \citetalias{hernandez07a}. In general, diskless stars have WH$_\alpha$\_10\% < 270 km $s^{-1}$
which is the accretion cutoff proposed by \citet{white03}.
We found 8 objects with infrared excess at 8{\micron} consistent with optically thick disks 
but that are not accreting or are accreting below the detectable level (hereafter
very slow accretor). These stars are SO 467, SO 738, SO 435, SO 674, 
SO 451, SO 247, SO 697 and SO 662. The star SO 823 was also included as a very slow 
accretor candidate. This star has a strong absorption component at H$\alpha$, 
we were unable to measure WH$_\alpha$\_10\% and thus it was not included in 
Figure \ref{f:EW10alfa}. On the other hand, low resolution spectra of 
SO 823 shows a single emission H$\alpha$ profile which suggests variability 
in the profile of H$\alpha$. Additionally, photometric 
measurements from CVSO indicate that SO 823 is a variable star with 
amplitude of $\Delta$V$\sim$2.5 magnitudes. Objects such as UX Ori type 
and AA Tau type show photometric and spectroscopic variability, 
in which, H$\alpha$ could change from single emission line profile to a 
profile with a strong central absorption component. 

Figure \ref{f:irac} shows the distribution on the IRAC color-color diagram ([5.8]-[8.0] vs. [3.6]-[4.5])
for stars studied in Figure\ref{f:EW10alfa} .
The very slow accretor candidates (including the star SO 823) were plotted with 
solid circles. Stars with optically thick disks classified as accretor 
and very slow accretor candidates fall in similar region in this plot. 
Thus, very slow accretors candidates have similar infrared excesses in the IRAC bands 
in comparison with stars with optically thick disks classified as accretors 
using the accretion cutoff proposed by \citet{white03}.

Low resolution spectroscopic analysis (\S \ref{sptclass}) supports 
the low accretion level for stars SO 823, SO 467, SO 435 and SO 451.
Moreover, in Figure \ref{f:CW_TTS} we can identify four additional 
very low accretor candidates (SO 566, SO 598, SO 682 and SO 967)
bearing optically thick disks which were classified as non-accretor 
following the criteria from \citet{barrado03}. High resolution studies 
are necessary to determine whether these stars are accreting or not 
\citep[e.g.][]{espaillat08,ingleby11}.

Figure \ref{f:sed_Ha} shows normalized SEDs and H$_\alpha$ profiles of 
the very slow accretor candidates detected in our high resolution spectra. 
For comparison, we show SEDs and profiles of a diskless star (SO 669) 
and two stars classified as accretor in Figure \ref{f:EW10alfa} (SO 397, SO 462). 
Particularly,  the star SO 462 has WH$_\alpha$\_10\% on the accretion cutoff defined by \citet{white03}.
The star SO 397 exhibit infrared excesses below the median SED for class II stars in the \SOri cluster.
High inclination of the disk (edge-on) or dust settling could be responsible for the relatively small
infrared excesses observed in this star.
The star SO 823 has a jump in the SED between the 2MASS and the IRAC measurements which could be
caused by variability or an unresolved binary.
Some stars like SO 467, SO 435, SO 451 and SO 462 exhibit asymmetries in the H$\alpha$ profile
that are characteristic of accreting disks.
The emission of H$\alpha$ could have a contribution from the stellar chromosphere or
the stars could have variable accretion rate, so the WH$_\alpha$\_10\% could vary and 
would not be a robust quantitative indicator of accretion \citep[e.g.,][]{nguyen09a,nguyen09b}. 
Thus, we need additional studies to understand whether these stars have 
stopped accreting, or if they are in a passive phase in which accretion is 
temporally stopped or if accretion is below the measurable levels in WH$_\alpha$\_10\%.
Previous studies have found very low accretors stars in other
young stellar populations. Studying the accretion properties in 
Chamaleon I and $\rho$ Oph, \citet{natta04} found a population 
of very low mass objects with evidence of disks but no detectable 
accretion activity estimated using several indicators of accretion. 
\citet{nguyen09a} also found a group of stars in Chamaeleon I 
($\sim$ 2Myr old) with excess at 8{\micron} and accretion rates 
below the measurable levels in WH$_\alpha$\_10\%. They also found 
that non-accreting objects with disks do not seem to exist 
in the Taurus-Auriga star forming region.

On the other hand, Figure \ref{f:EW10alfa} shows two diskless 
stars (SO 616 and SO 229) exhibiting WH$_\alpha$\_10\% larger than the 
accretors limits. One of these stars, SO 229, was 
identified as a double lined spectroscopic binary from the cross 
correlation function while the star SO 616 was recognized as
possible binary based on radial velocity variability. 
These two stars have \ion{Li}{1}$\lambda$6708 in absorption and spectral types of G2.5 and K7,  
respectively (see \S \ref{sptclass}). In Figure \ref{f:rotator}, 
we compare the Hectochelle spectra of SO 229 and SO 616 to stars with similar spectral
types (G3 for SO 229 and K7 for SO 616). It is apparent that SO 299 and SO 616 have wider 
photospheric features. It is possible that wider spectral features are combined features 
from component in binary stars with similar  spectral types. 
Another possibility is that the broadening of spectral lines in these objects can 
be caused by fast rotation. Non-accreting objects with projected rotational velocity 
larger than $\sim$ 50 km $s^{-1}$ can generate WH$_\alpha$\_10\% > 270 km $s^{-1}$ 
\citep[see Figure 5 of ][]{jayawardhana06}.  A multi-epoch spectroscopic analysis 
will help to reveal the nature of these objects.

Low resolution spectroscopic analysis(\S \ref{sptclass}) shows
three diskless stars (SO 229, SO 1123 and SO 1368) that mimic
the H$\alpha$ width expected in stars with accretion disks (see Figure \ref{f:CW_TTS}). 
The star SO 299 shows broad photospheric features that could be produced 
by binarity or fast rotation (Figure \ref{f:rotator}).  
The star SO1123 has been classified as spectroscopic binary \citep{caballero07} 
and fast rotator with a very high projected rotational velocity \citep{alcala00}. 
The star SO 1368 is a variable star \citep{mayne07} with spectral type K6.5
and still does not have information about multiplicity or rotation. 
These diskless objects with H$\alpha$ in emission could have substantial 
contribution produced by strong chromospheric activity  
related to fast rotation.

\section{Summary and Conclusions}
\label{sec:conc}

Combining 2MASS data \citep{cutri03}, new optical photometry obtained with the OSMOS instrument, 
an updated optical photometry from the CIDA Variability Survey of Orion and photometric information from 
previous work \citep{sherry04, kenyon05, mayne07, kharchenko09}, we defined a list of photometric 
candidates of the \SOri cluster. Substantial contamination is expected for solar type stars 
(V-J$\sim$1.5) because a branch of field stars cross the PMS population of the cluster in the 
color magnitude diagram used to select photometric candidates. A subset of these candidates were 
characterized using spectroscopic data. 

We have applied a consistent spectral classification scheme aimed at PMS stars. 
Low resolution spectroscopic data set for spectral typing come from several instruments 
with similar spectral coverage and resolution. We were able to determine spectral types 
for 340 objects located in the general region of the \SOri cluster. Analysis of this data set
enables us to define membership indicators based on the accretion status obtained 
from H$\alpha$ line \citep{barrado03}, the presence of \ion{Li}{1}$\lambda$6708 in absorption 
(for LMS and for VLMS) and for a subset of VLMS observed with the Hectospec instrument the 
intensity of the line \ion{NaI}{1}$\lambda8195$. So far, this analysis constitutes the largest 
homogeneous spectroscopic characterization of members in the {\SOri.} Additionally, we were able 
to determine radial velocities for 95 stars out of a total sample of 142 objects observed at 
high resolution in the general region of the cluster. For this high resolution spectroscopic 
data set we also determined membership indicators based on the presence of \ion{Li}{1}$\lambda$6708 
in absorption and the width of the H$\alpha$ line \citep{white03}. The radial velocity distribution 
for members of the cluster is in agreement with previous works \citep[e.g.][]{sacco08,maxted08,gonzalez08}. 

We have identified and assigned spectral types to 178 bona fide members of the \SOri cluster, 
combining results from our spectroscopic analysis, previous membership (based on radial velocity distribution 
and the presence  of \ion{Li}{1}$\lambda$6708) and other membership indicators like 
detection of protoplanetary disks, X ray emission, proper motion criteria, distance from
the central star, variability and photometric membership probability.  We also have identified
14 bona fide members of the cluster using high resolution analysis and reported membership 
indicators. Additionally, we have identified and assigned spectral types to 25 probable members 
of the cluster. Finally 36 stars do not have enough information to assign membership 
(uncertain members).

Of particular interest are stars bearing an optically thick disks and 
narrow H$\alpha$ profile not expected in stars with accretion disks. 
Also objects with no infrared excess and H$\alpha$ line that
mimic the H$\alpha$ width expected in stars with accretion disks 
(maybe because binarity or fast rotation). Additional multi-epoch observations
are necessary to reveal the nature of these objects.

Additional information, the optical spectra, UBVRI magnitudes, 
membership indicators, and spectral energy distribution for each object 
analyzed in this work are reported on the World Wide Web 
\footnote{http://sigmaori.cida.gob.ve/}.
Information in this website is a work in progress and we expect to 
increase the spectroscopic data base and the membership analysis
for additional stars in the \SOri cluster.  

\acknowledgements

We thank John Tobin for his advice during the reduction 
and analysis of the Hectochelle data. Thanks 
to Gail Schaefer and Jonh Monnier for insightful
communications about the distance of the \SOri cluster.
{ An anonymous referee provided many insightful comments.}
We thank Cecilia Mateu for providing the updated version 
of the CVSO catalog. We also thank Nelson Caldwell, Andy Szentgyorgyi,
Perry Berlind, Michael Calkins and Susan Tokarz for 
their help in obtaining and processing the Hectospec 
and FAST spectra. I would like to thank the institutions and personnel that support data 
acquisition at the Observatorio Astron\'{o}mico Nacional de Llano del Hato (CIDA),
the Observatorio Astron\'{o}mico Nacional at San Pedro Martir (UNAM),
the Observatorio Astrof\'{\i}sico Guillermo Haro (INAOE),
and the MDM Observatory (University of Michigan).
This publication makes use of data products
from the CIDA Equatorial Variability Survey, obtained with the
J. Stock telescope at the Venezuela National Astronomical Observatory, 
which is operated by CIDA for the Ministerio del Poder Popular para Ciencia, Tecnolog\'{\i}a 
e Innovaci\'{o}n of Venezuela.
We also make use data products from UKIDSS (obtained as part of 
the UKIRT Infrared Deep Sky Survey)  and from the Two Micron
All Sky Survey, which is a joint project of the University of
Massachusetts and the Infrared Processing and Analysis Center/
California Institute of Technology.
This work makes use of observations made with the Spitzer Space 
Telescope (GO-1 0037 and GO-1 0058), which is operated by the Jet 
Propulsion Laboratory, California Institute of Technology, under 
a contract with NASA.
Support for this work was provided by CIDA, the University of Michigan
and by the grant UNAM-DGAPA IN109311.

\appendix
\section{Known members, new disk bearing candidates and photometric candidates not listed in Table \ref{t:mem1} and Table \ref{t:mem2}}
\label{app1}

Table \ref{t:appA} includes known members compiled in \S \ref{known members} 
and new disk bearing candidates identified in \S{\ref{s:mips} which 
were not studied in \S\ref{sptclass} and \S\ref{radialvelocity}.
Except for stars observed using the Hectospec and Hectochelle multifibers 
spectrographs, our target selection includes stars with V$<$16.5 (M$_* >$0.35\msun; see \S\ref{s:spec}). 
Out of 125 stars included in Table \ref{t:appA}, 107 stars (85.6\%) are fainter 
than our target selection limit. Out of 18 stars with V<16.5, 3 stars are new disk 
bearing candidates. Sorted by optical brightness, Table \ref{t:appA} shows in
columns (1) and (2) source designations from \citet{hernandez07a} and \citet{cutri03}, respectively.
Column (3) shows visual magnitudes (\S\ref{sec:phot}). Membership information based on 
the presence of protoplanetary disks, the presence of \ion{Li}{1}$\lambda$6708, 
radial velocity measurements, X ray emission, proper motion, distance from the center of the cluster
and variability are in columns (4), (5), (6), (7), (8), (9) and (10), respectively.
Column (11) gives the photometric membership probability calculated in \S\ref{s:phot_prob}.
Spectral types and references are in columns (11) and (12), respectively. 

Sorted by optical brightness, Table \ref{t:appB} includes additional photometric candidates 
with photometric membership probability higher than $\sim$32\% (within the dashed lines in 
Figure \ref{f:CMD_rot}). In this Tables we also include photometric candidates selected by 
\citet[][;UKIDSS'candidates]{lodieu09} which are brighter than the completeness limit of 
2MASS catalog (J$\sim$15.8). Out of 11  UKIDSS'candidates not included in 
Table \ref{t:mem1}, 4 UKIDSS'candidates were included in Table \ref{t:appA}. 
The remaining 7 stars are listed in Table \ref{t:appB}.

\begin{deluxetable}{cccccccc}
\tabletypesize{\scriptsize}
\tablewidth{0pt}
\tablecaption{Compiled membership information of the \SOri cluster \label{t:knownmembers}}
\tablehead{
\colhead{Name} & \colhead{2massID} & \colhead{RA} & \colhead{DEC}      & \colhead{Disk}  & \colhead{\ion{Li}{1}} & \colhead{RV} & \colhead{X Ray}\\
\colhead{H07} & \colhead{  }     & \colhead{degree}   & \colhead{degree } & \colhead{type} & \colhead{Ref} & \colhead{Ref} & \colhead{Ref} \\
}
\startdata
SO27 & 05372306-0232465 & 84.346086 & -2.546276 & III & \nodata & \nodata & 19 \\
SO59 & 05372806-0236065 & 84.366955 & -2.60182 & III & \nodata & \nodata & 16,19 \\
SO60 & 05372831-0224182 & 84.367989 & -2.40506 & III & \nodata & \nodata & 16,19 \\
SO73 & 05373094-0223427 & 84.378949 & -2.395214 & II & \nodata & \nodata & 19 \\
SO77 & 05373153-0224269 & 84.381401 & -2.407489 & III & 6 & \nodata & 16,19 \\
SO107 & 05373514-0226577 & 84.396435 & -2.44937 & III & \nodata & \nodata & 19 \\
SO116 & 05373648-0241567 & 84.40202 & -2.699093 & III & 3,5,12 & \nodata & \nodata \\
SO762 & 05385060-0242429 & 84.710859 & -2.711918 & II & 5 & 5,13 & \nodata \\
SO765 & 05385077-0236267 & 84.711581 & -2.607432 & III & 5,8 & 5,8 & 18 \\
SO976 & 05391699-0241171 & 84.820833 & -2.688091 & III & \nodata & \nodata & 14,18,19 \\
SO1185 & 05394340-0253230 & 84.930841 & -2.889736 & III & \nodata & 5 & \nodata \\
\nodata & 05381741-0240242 & 84.572574 & -2.673413 & \nodata & 2,3,5 & 13 & \nodata \\
\nodata & 05382557-0248370 & 84.606573 & -2.810284 & \nodata & 2,3 & \nodata & \nodata \\
\nodata\tablenotemark{1} & 05390756-0212145 & 84.78150  & -2.204030 & \nodata & \nodata & \nodata & \nodata \\
\nodata & 05390893-0257049 & 84.787241 & -2.951387 & \nodata & 6 & \nodata & \nodata \\
\nodata & 05402018-0226082 & 85.084123 & -2.435633 & \nodata & \nodata & \nodata & 14 \\
\nodata & 05402076-0230299 & 85.086534 & -2.508313 & \nodata & \nodata & \nodata & 14 \\
\nodata & 05402256-0233469 & 85.094019 & -2.56303 & \nodata & 6 & 7 & 14 \\
\nodata & 05402301-0236100 & 85.095882 & -2.6028 & \nodata & 6 & 13 & \nodata \\
\enddata
\tablecomments{Table \ref{t:knownmembers} is published in its entirety in the electronic edition of
the {\it Astrophysical Journal}. A portion is shown here for guidance
regarding its form and content.}
\tablenotetext{~}{References: (1) \citet{wolk96}; (2) \citet{zapatero02}; (3) \citet{barrado03}; (4) \citet{andrews04};
(5) \citet{kenyon05}; (6) \citet{caballero06}; (7) \citet{gonzalez08}; (8) \citet{sacco08}; (9) \citet{caballero12};
(10) \citet{alcala96}; (11) \citet{caballero06b}; (12) \citet{muzerolle03}; (13) \citep{maxted08};
(14) \citet{caballero08a}; (15) \citet{skinner08}; (16) \citet{lopez08}; (17) \citet{caballero09}; 
(18) \citet{caballero10}; (19) XMMN-SSC,2010); (20) \citet{burningham05}}
\tablenotetext{~}{Disk type: III=diskless, II=thick disk, I=class I, EV= evolved disk, TD=transition disk, DD=Debris disk}
\tablenotetext{1}{Strong H$\alpha$ line and Accretion disk \citep{scholz04,scholz09} }
\tablenotetext{2}{Strong H$\alpha$ line and Accretion disk \citep{caballero08} }
\end{deluxetable}

\begin{deluxetable}{ccccccccccc}
\rotate
\tabletypesize{\scriptsize}
\tablewidth{0pt}
\tablecaption{Photometric candidates \label{t:photcan}}
\tablehead{
\colhead{Name} & \colhead{2massID} & \colhead{RAJ2000} &  \colhead{DECJ2000} & \colhead{U} & \colhead{B}  & \colhead{V} &\colhead{R} &  \colhead{I} &\colhead{J} &\colhead{Ref} \\
\colhead{H07}    & \colhead{  }    & \colhead{deg} & \colhead{deg}   & \colhead{mag}  & \colhead{mag} & \colhead{mag} & \colhead{mag} & \colhead{mag} & \colhead{mag} & \colhead{ }\\
}
\startdata
\nodata & 05371360-0229293 & 84.30668& -2.49147 & \nodata & 16.48$\pm$0.02 & 15.19$\pm$0.00 & \nodata & 13.52$\pm$0.01 & 12.25$\pm$0.04 & CLC \\
\nodata & 05371373-0258190\tablenotemark{3} & 84.30724& -2.97196 & \nodata & 21.53$\pm$0.64 & 21.08$\pm$0.02 & 19.74$\pm$0.02 & 17.98$\pm$0.05 & 16.11$\pm$0.11 & CLC \\
\nodata & 05371407-0221089 & 84.30863& -2.35249 & \nodata & \nodata & 13.39$\pm$0.02 & \nodata & \nodata & 11.18$\pm$0.02 & CVSO \\
\nodata & 05372254-0259363 & 84.34395& -2.99343 & \nodata & \nodata & 15.06$\pm$0.01 & 14.13$\pm$0.01 & 13.15$\pm$0.01 & 11.73$\pm$0.02 & S04 \\
SO77 & 05373153-0224269 & 84.38140& -2.40749 & \nodata & 17.32$\pm$0.02 & 15.85$\pm$0.01 & 14.72$\pm$0.01 & 13.45$\pm$0.01 & 12.11$\pm$0.03 & S04 \\
SO189 & 05374924-0253118 & 84.45518& -2.88662 & \nodata & \nodata & 18.04$\pm$0.07 & 16.94$\pm$0.02 & \nodata & 14.44$\pm$0.04 & CVSO \\
SO199 & 05374981-0246032 & 84.45757& -2.76756 & \nodata & \nodata & 13.74$\pm$0.02 & \nodata & \nodata & 11.69$\pm$0.03 & CVSO \\
SO247 & 05375486-0241092 & 84.47860& -2.68590 & \nodata & 19.68$\pm$0.18 & 18.34$\pm$0.09 & 17.14$\pm$0.06 & 15.43$\pm$0.03 & 13.50$\pm$0.03 & S04 \\
SO389 & 05381223-0218124 & 84.55097& -2.30345 & \nodata & 16.79$\pm$0.02 & 15.39$\pm$0.01 & 14.51$\pm$0.01 & 13.71$\pm$0.01 & 12.66$\pm$0.00\tablenotemark{1} & CLC \\
SO411 & 05381412-0215597 & 84.55884& -2.26660 & \nodata & 10.93$\pm$0.06 & 10.45$\pm$0.06 & \nodata & \nodata & 9.39$\pm$0.02 & K09 \\
SO563 & 05383157-0235148 & 84.63158& -2.58746 & 18.19$\pm$0.18 & 17.17$\pm$0.02 & 15.85$\pm$0.01 & 14.88$\pm$0.01 & 13.77$\pm$0.01 & 11.52$\pm$0.03 & OS \\
SO567 & 05383243-0251215 & 84.63514& -2.85597 & \nodata & 14.20$\pm$0.01 & 13.34$\pm$0.01 & \nodata & 12.24$\pm$0.01 & 11.60$\pm$0.03 & CLC \\
SO706 & 05384480-0233576 & 84.68668& -2.56601 & 14.82$\pm$0.02 & 13.73$\pm$0.00 & 12.49$\pm$0.00 & 11.77$\pm$0.00 & 11.11$\pm$0.00 & 10.01$\pm$0.03 & OS \\
\nodata & 05384503-0258340 & 84.68766& -2.97612 & \nodata & 21.78$\pm$0.84 & 19.61$\pm$0.01 & 18.24$\pm$0.01 & 16.75$\pm$0.01 & 15.20$\pm$0.04 & CLC \\
SO739 & 05384818-0244007 & 84.70078& -2.73355 & 21.27$\pm$0.49 & 21.43$\pm$0.09 & 19.76$\pm$0.02 & 18.11$\pm$0.01 & 16.13$\pm$0.01 & 14.07$\pm$0.03 & OS \\
SO927 & 05391151-0231065 & 84.79798& -2.51849 & 16.75$\pm$0.07 & 16.98$\pm$0.02 & 15.66$\pm$0.01 & 14.64$\pm$0.01 & 13.56$\pm$0.01 & 11.99$\pm$0.03 & OS \\
SO929 & 05391163-0236028 & 84.79846& -2.60080 & 17.04$\pm$0.07 & 15.82$\pm$0.01 & 14.47$\pm$0.00 & 13.61$\pm$0.00 & 12.75$\pm$0.00 & 11.62$\pm$0.03 & OS \\
SO947 & 05391453-0219367 & 84.81056& -2.32687 & \nodata & 17.33$\pm$0.02 & 15.90$\pm$0.02 & 14.83$\pm$0.02 & 13.60$\pm$0.02 & 12.16$\pm$0.03 & S04 \\
SO1081 & 05393056-0238270\tablenotemark{3} & 84.87733& -2.64084 & \nodata & 19.36$\pm$0.13 & 17.82$\pm$0.05 & 16.61$\pm$0.03 & 15.21$\pm$0.02 & 13.81$\pm$0.03 & S04 \\
SO1224 & 05394891-0229110\tablenotemark{3} & 84.95381& -2.48641 & \nodata & 19.02$\pm$0.10 & 17.34$\pm$0.03 & 16.24$\pm$0.02 & 14.70$\pm$0.01 & 13.28$\pm$0.03 & S04 \\
\nodata & 05400405-0255375 & 85.01689& -2.92711 & \nodata & \nodata & 17.46$\pm$0.05 & 16.40$\pm$0.02 & 15.18$\pm$0.02 & 14.07$\pm$0.03 & CVSO \\
\nodata & 05402378-0228261 & 85.09912& -2.47394 & \nodata & \nodata & 17.78$\pm$0.05 & 16.74$\pm$0.02 & \nodata & 13.86$\pm$0.04 & CVSO \\
\enddata
\tablecomments{Table \ref{t:photcan} is published in its entirety in the electronic edition of the {\it Astrophysical Journal}. 
A portion is shown here for guidance regarding its form and content.}
\tablenotetext{~}{References: OS OSMOS (\S\ref{osmos}); CVSO CIDA Variability Survey of Orion \citep{briceno05}; S04 \citet{sherry04}; CLC Cluster Collaboration \citep{kenyon05,mayne07}; 
K09 \citet{kharchenko09}}
\tablenotetext{1}{J magnitude from UKIDSS}
\tablenotetext{2}{Known members not selected as photometric candidates}
\tablenotetext{3}{Sources labeled as galaxies by \citet{lawrance13}}
\end{deluxetable}

\begin{deluxetable}{cccccc}
\tabletypesize{\scriptsize}
\tablewidth{0pt}
\tablecaption{Low Resolution Spectrographs \label{t:spectrograph}}
\tablehead{
\colhead{Observatory} & \colhead{Telescope} & \colhead{Spectrograph} & \colhead{Resolution at H$\alpha$} & \colhead{REF} & \colhead{Spectral Coverage}\\
\colhead{ } & \colhead{ meters}    & \colhead{ }   & \colhead{ \AA } & \colhead{ }       & \colhead{\AA} \\
}
\startdata
MMTO             & 6.5 & Hectospec (H)        & 6  & \citet{fabricant05}     & 3650-9200\\ 
FLWO             & 1.5 & FAST      (F)       & 6  & \citet{fabricant98}     & 3800-7200 \\
MDMO             & 1.3 & OSU-CCDS  (O)       & 6.5  & Measured with arc lines & 3900-7300  \\
OAN-SPM          & 2.1 & Boller \& Chiven (S) & 5.5  & Measured with arc lines & 3900-7200\\
Guillermo-Haro   & 2.1 & Boller \& Chiven (C) & 10 & Measured with arc lines & 4100-7300 \\
\enddata
\end{deluxetable}

\begin{deluxetable}{cccccccccccc}
\tabletypesize{\scriptsize}
\tablewidth{0pt}
\tablecaption{Low resolution analysis \label{t:lowres}}
\tablehead{
\colhead{Name} & \colhead{2massID} & \colhead{INST} & \colhead{Spectral} &  \colhead{\ion{Li}{1}} &\colhead{\ion{Li}{1}}  & \colhead{H$\alpha$} & \colhead{H$\alpha$} & \colhead{\ion{Na}{1}} & \colhead{\ion{Na}{1}} & \colhead{Av$_{KH95}$} & \colhead{Av$_{PM13}$} \\
\colhead{H07}    & \colhead{  } & \colhead{  }        & \colhead{Types}    & \colhead{\AA}  & \colhead{flag}                 & \colhead{ \AA}      &  \colhead{flag}   &   \colhead{\AA }       &  \colhead{flag}       &\colhead{mag}        & \colhead{mag} \\
}
\startdata
\nodata & 05384476-0236001 & F & B0.0$\pm$1.5 & 0.0 & 0 & 3.1 & E$_{abs}$ & \nodata & \nodata & 0.00 & 0.00 \\
\nodata & 05384561-0235588 & F & B2.0$\pm$1.5 & 0.0 & 0 & 4.3 & E$_{abs}$ & \nodata & \nodata & 0.10 & 0.00 \\
\nodata & 05402018-0226082 & F & B4.0$\pm$1.5 & 0.0 & 0 & 5.0 & E$_{abs}$ & \nodata & \nodata & 0.17 & 0.00 \\
SO139 & 05374047-0226367 & F & A3.5$\pm$2.5 & 0.0 & 0 & 11.0 & E$_{abs}$ & \nodata & \nodata & 0.00 & 0.14 \\
SO521 & 05382752-0243325 & F & A8.0$\pm$2.5 & 0.0 & 0 & 8.3 & E$_{abs}$ & \nodata & \nodata & 0.18 & 0.84 \\
SO338 & 05380649-0228494 & S & F3.5$\pm$2.0 & 0.1 & 1 & 3.5 & E$_{abs}$ & \nodata & \nodata & 0.72 & 0.72 \\
SO1352 & 05400696-0228300 & F & F7.5$\pm$2.0 & 0.0 & 0 & 3.6 & E$_{abs}$ & \nodata & \nodata & 1.36 & 1.44 \\
SO1246 & 05395118-0222461 & H & G1.0$\pm$2.5 & 0.0 & 0 & 1.5 & E$_{abs}$ & \nodata & \nodata & 2.41 & 2.49 \\
SO550 & 05383008-0221198 & H & G1.5$\pm$2.5 & 0.0 & 0 & 1.8 & E$_{abs}$ & \nodata & \nodata & 2.05 & 2.12 \\
SO1286 & 05395658-0246236 & H & G5.0$\pm$2.5 & 0.0 & 0 & 2.1 & nAcr & \nodata & \nodata & 1.83 & 1.70 \\
SO29 & 05372330-0229133 & H & G9.5$\pm$2.0 & 0.0 & 0 & 1.8 & nAcr & \nodata & \nodata & 0.51 & 0.27 \\
SO449 & 05381879-0217138 & H & K0.0$\pm$3.0 & 0.0 & 0 & 0.0 & nAcr & \nodata & \nodata & 1.00 & 0.83 \\
SO903 & 05390828-0249462 & F & K5.5$\pm$1.5 & 0.0 & 0 & 0.4 & nAcr & \nodata & \nodata & 0.00 & 0.00 \\
SO1113 & 05393511-0247299 & F & K5.5$\pm$1.5 & 0.5 & 2 & -1.1 & nAcr & \nodata & \nodata & 0.41 & 0.00 \\
SO1274 & 05395465-0246341 & S & K6.0$\pm$1.0 & 0.3 & 2 & -36.5 & Acr & \nodata & \nodata & 1.40 & 0.92 \\
SO611 & 05383546-0231516 & H & K7.0$\pm$1.0 & 0.6 & 2 & -2.7 & Acr? & \nodata & \nodata & 0.48 & 0.11 \\
SO35 & 05372384-0248532 & H\tablenotemark{1} & M0.0$\pm$1.5 & 0.0 & 1 & 2.4 & nAcr & \nodata & \nodata & 0.77 & 0.46\\
SO726 & 05384746-0235252 & H & M0.5$\pm$1.0 & 0.4 & 2 & -23.0 & Acr & 0.94$\pm$0.08 & \nodata & 0.03 & 0.00 \\
SO682\tablenotemark{2} & 05384227-0237147 & C & M0.5$\pm$1.0 & 0.2 & 1 & -2.7 & nAcr & \nodata & \nodata & 0.67 & 0.34 \\
SO600 & 05383479-0239300 & F & M1.0$\pm$0.5 & 0.0 & 0 & 0.9 & nAcr & \nodata & \nodata & 0.71 & 0.34 \\
SO1282 & 05395594-0220366 & H & M2.0$\pm$0.5 & 0.4 & 2 & -8.2 & nAcr & 1.37$\pm$0.10 & \nodata & 0.62 & 0.23 \\
SO444 & 05381824-0248143 & H & M3.0$\pm$0.5 & 0.5 & 2 & -3.8 & nAcr & 0.83$\pm$0.08 & Y & 0.12 & 0.08 \\
SO562 & 05383141-0236338 & H & M3.5$\pm$1.5 & 0.1 & 2 & -77.7 & Acr & 0.52$\pm$0.10 & Y & 0.07 & 0.14 \\
SO1053 & 05392650-0252152 & H & M4.0$\pm$0.5 & 0.4 & 2 & -3.6 & nAcr & 1.02$\pm$0.09 & Y & 0.00 & 0.00 \\
SO300 & 05380107-0245379 & H & M4.5$\pm$0.5 & 0.4 & 2 & -92.6 & Acr & 0.65$\pm$0.10 & Y & 0.14 & 0.26 \\
SO460 & 05382021-0238016 & H & M5.0$\pm$0.5 & 0.4 & 2 & -10.6 & nAcr & 1.44$\pm$0.14 & Y & 0.00 & 0.00 \\
SO283 & 05375840-0241262 & H & M5.5$\pm$0.5 & 0.6 & 1 & -14.3 & Acr? & 0.99$\pm$0.13 & Y & 0.00 & 0.00 \\
SO457 & 05381975-0236391 & F & M6.0$\pm$0.5 & 0.0 & 0 & 0.2 & nAcr & \nodata & \nodata & 0.00 & 0.32 \\
SO209 & 05375110-0226074 & H\tablenotemark{1} & M6.5$\pm$2.5 & 0.0 & 1 & -13.3 & nAcr & \nodata & \nodata & \nodata & \nodata \\
\enddata
\tablecomments{Table \ref{t:lowres} is published in its entirety in the electronic edition of
the {\it Astrophysical Journal}. A portion is shown here for guidance regarding its form and content.}
\tablenotetext{~}{INST: See Table \ref{t:spectrograph}}
\tablenotetext{~}{\ion{Li}{1} flag: (2) \ion{Li}{1} is present in absoprtion; (1) \ion{Li}{1} uncertain; 
(0) \ion{Li}{1} is not present on the spectrum}
\tablenotetext{~}{H$\alpha$ flag: (Acr) accretor; (nAcr) non-accretor; (Acr?) uncertain; 
(E$_{abs}$) star ealier than G5 with H$\alpha$ in absorption; (ET$_{em}$) star ealier than G5 with H$\alpha$ in emission}
\tablenotetext{~}{\ion{NaI}{1} flag: (Y) member; (N) non-member}
\tablenotetext{1}{Low signal to noise (SN$\lesssim$15)}
\tablenotetext{2}{Slow accretor candidate}
\tablenotetext{3}{Diskless stars that mimic the H$\alpha$ width expected in stars with accretion disks} 

\end{deluxetable}


\begin{deluxetable}{ccccccccc}
\tabletypesize{\scriptsize}
\tablewidth{0pt}
\tablecaption{High resolution analysis \label{t:hectochelle}}
\tablehead{
\colhead{Name} & \colhead{2massID} & \colhead{RV} &  \colhead{RV} & \colhead{\ion{Li}{1}} & \colhead{W10\_H$\alpha$}  & \colhead{Acretor} &\colhead{Disk} & \colhead{Comments} \\
\colhead{H07}    & \colhead{  } & \colhead{km/s} & \colhead{flag}   & \colhead{\AA}  & \colhead{km/s} & \colhead{ } & \colhead{type} & \colhead{ }\\
}
\startdata
SO52  &  05372692-0221541  &  9.6$\pm$0.3  & 0  &  \nodata  &  \nodata & \nodata &  III & H$\alpha$ in absorption, SB2\\
SO59  &  05372806-0236065  &  29.8$\pm$0.9  & 2  &  \nodata  &  115.203 & N &  III & \\
SO74  &  05373105-0231436  &  \nodata  & \nodata  &  \nodata  &  \nodata & \nodata &  III      & H$\alpha$ in absorption with central emission \\
SO247  &  05375486-0241092  &  \nodata  & \nodata  &  0.25$\pm$0.12  &  192.322 & N &  II  & \\ 
SO300  &  05380107-0245379  &  \nodata  & \nodata  &  \nodata  &  413.678 & Y &  II         & \\
SO341  &  05380674-0230227  &  33.7$\pm$0.6  & 2  &  0.59$\pm$0.07  &  427.684 & Y &  II & \\
SO374  &  05380994-0251377  &  30.0$\pm$0.6  & 2  &  0.44$\pm$0.06  &  411.430 & Y &  II & \\
SO411  &  05381412-0215597  &  22.2$\pm$1.1  & 1  &  0.16$\pm$0.01  &  294.276 & Y &  TD & \\
SO489  &  05382354-0241317  &  \nodata  & \nodata  &  0.41$\pm$0.17  &  143.305 & N &  III         & \\
SO514  &  05382684-0238460  &  \nodata  & \nodata  &  \nodata  &  \nodata         & \nodata &  II & inverse P Cygni profile \\
SO518  &  05382725-0245096  &  28.8$\pm$0.6  & 2  &  0.43$\pm$0.02  &  575.190 & Y &  II &  \\
SO539  &  05382911-0236026  &  33.2$\pm$0.7  & 2  &  0.77$\pm$0.10  &  152.280 & N &  III & \\
SO548  &  05382995-0215405  &  \nodata  & \nodata  &  \nodata  &  133.636 & N &  III         & \\
SO616  &  05383587-0230433  &  23.8$\pm$5.6  & 1  &  0.82$\pm$0.03  &  465.422 & Y &  III & RV$_{VAR}$\\ 
SO697  &  05384423-0240197  &  29.4$\pm$0.2  & 2  &  0.56$\pm$0.02  &  198.812 & N &  II & \\ 
SO752  &  05384945-0249568  &  -49.2$\pm$0.6  & 0  &  \nodata  &  \nodata & \nodata &  III & Double central absorption \\
SO1097  &  05393291-0247492  &  27.8$\pm$1.3  & 2  &  0.59$\pm$0.03  &  168.462 & N &  III & \\
SO1251  &  05395253-0243223  &  46.0$\pm$0.2  & 0  &  0.08$\pm$0.04  &  \nodata & \nodata &  III & H$\alpha$ in absorption, EW[\ion{Li}{1}]$<$0.1\\
SO1352  &  05400696-0228300  &  34.9$\pm$0.2  & 2  &  \nodata  &  \nodata & \nodata &  III & H$\alpha$ in absorption \\
SO1361  &  05400889-0233336  &  30.1$\pm$0.6  & 2  &  0.48$\pm$0.02  &  362.225 & Y &  II & \\
SO1370  &  05401304-0228314  &  \nodata  & \nodata  &  \nodata                &  \nodata & \nodata &  III & H$\alpha$ in absorption\\
\enddata
\tablecomments{Table \ref{t:hectochelle} is published in its entirety in the electronic edition of
the {\it Astrophysical Journal}. A portion is shown here for guidance regarding its form and content.}
\tablenotetext{~}{RV flag: (2) kinematic member; (1) sparser stellar population candidate; (0) kinematic non-member}
\tablenotetext{~}{Accretor flag: (Y) accretor (W10\_H$\alpha >$270 km/s); (N) non-accretor (W10\_H$\alpha <$270 km/s)}
\tablenotetext{~}{Disk type: III=diskless, II=thick disk, I=class I, EV= evolved disk, TD=transition disk, DD=Debris disk}
\tablenotetext{~}{RV$_{var}$: binary candidate by radial velocity variability}
\tablenotetext{~}{SB2: binary candidate identified in the cross correlation function}
\end{deluxetable}


\begin{deluxetable}{ccccccccccccccccc}
\rotate
\tabletypesize{\scriptsize}
\tablewidth{0pt}
\tablecaption{Membership for stars with spectral types \label{t:mem1}}
\tablehead{
\colhead{Name} & \colhead{2massID} & \colhead{Spectral} & \colhead{\ion{Li}{1} flag} &  \colhead{\ion{Li}{1}} &\colhead{RV}  & \colhead{RV} & \colhead{H$\alpha$ flag} & \colhead{NaI} & \colhead{Disk} & \colhead{Xray} & \colhead{PM} & \colhead{Dist} & \colhead{Var} & \colhead{\%$_{pho}$}  & \colhead{A$_V$} & \colhead{Member}\\
\colhead{H07}    & \colhead{  } & \colhead{types}        & \colhead{Low High}    & \colhead{Ref}  & \colhead{flag}  & \colhead{Ref}  &  \colhead{Low High}   &   \colhead{flag} &  \colhead{type}       &\colhead{Ref}        & \colhead{flag} & \colhead{flag} & \colhead{Ref} & \colhead{ } & \colhead{}& \colhead{flag} \\
}
\startdata
\nodata & 05384476-0236001 & B0.0$\pm$1.5 & 0 \nodata &	\nodata	& \nodata &	\nodata	& E$_{abs}$ \nodata & \nodata & \nodata &	14,15,17,18,19	& 2 & 1 &	\nodata	& 0.1 & 0.00 &  M:\tablenotemark{1} \\
SO139 & 05374047-0226367 & A3.5$\pm$2.5 & 0 \nodata &	\nodata	& \nodata &	\nodata	& E$_{abs}$ \nodata & \nodata & \nodata &	\nodata	& 2 & 1 &	\nodata	& 86.1 & 0.14 & U:\tablenotemark{3} \\
SO411 & 05381412-0215597 & F7.5$\pm$2.5 & 2 2 &	\nodata	& 1 &	\nodata	& E$_{em}$ Y & \nodata & TD &	14	& 2 & 0 &	\nodata	& 95.1 & 0.07 &  M:\\
SO37 & 05372414-0225520 & G0.5$\pm$2.0 & 0 \nodata &	\nodata	& \nodata &	\nodata	& E$_{abs}$ \nodata & \nodata & \nodata &	\nodata	& 2 & 0 &	\nodata	& \nodata & \nodata & U: \\
SO1307 & 05395930-0222543 & G2.0$\pm$2.5 & 2 \nodata &	\nodata	& \nodata &	\nodata	& E$_{abs}$ \nodata & \nodata & III &	14	& 2 & 0 &	\nodata	& 44.9 & 0.63 &  P: \\
SO981 & 05391807-0229284 & G7.5$\pm$2.5 & 1 \nodata &	\nodata	& \nodata &	\nodata	& nAcr \nodata & \nodata & DD &	14,17,18,19	& 2 & 1 &	\nodata	& 52.4 & 0.56 & M: \\
\nodata & 05392639-0215034 & K4.5$\pm$2.0 & 2 \nodata &	6	& \nodata &	\nodata	& Acr \nodata & \nodata & II &	\nodata	& 2 & 0 &	21,22,24	& 32.7 & 0.36 & M:\tablenotemark{19} \\
SO670 & 05384135-0236444 & M2.0$\pm$1.0 & 1 \nodata &	8	& \nodata &	8,13	& \nodata \nodata & \nodata & III &	15,18,19	& \nodata & 1 &	\nodata	& 42.3 & 0.30 & M: \\
SO27 & 05372306-0232465 & M3.0$\pm$1.0 & 1 \nodata &	\nodata	& \nodata &	\nodata	& nAcr \nodata & Y & III &	19	& \nodata & 0 &	\nodata	& 7.4 & 0.57 & P: \\
SO545 & 05382961-0225141 & M4.0$\pm$1.5 & 1 \nodata &	5,12	& \nodata &	\nodata,5	& \nodata \nodata & \nodata & III &	\nodata	& \nodata & 1 &	24	& 79.3 & 2.21 &  M: \\
SO466 & 05382089-0251280 & M5.5$\pm$2.0 & 1 \nodata &	\nodata	& \nodata &	13	& Acr? \nodata & \nodata & III &	\nodata	& \nodata & 1 &	\nodata	& 87.1 & 0.61 & M: \\
SO457 & 05381975-0236391 & M6.0$\pm$0.5 & 0 0 &	\nodata	& \nodata &	\nodata	& nAcr Y & \nodata & I &	\nodata	& 2 & 1 &	22	& 0.0 & 0.32 &  N:\tablenotemark{22} \\
SO209 & 05375110-0226074 & M6.5$\pm$2.5 & 1 \nodata &	\nodata	& \nodata &	\nodata	& nAcr \nodata & \nodata & III &	\nodata	& \nodata & 1 &	\nodata	& \nodata & \nodata & U:\\
\enddata
\tablecomments{Table \ref{t:mem1} is published in its entirety in the electronic edition of
the {\it Astrophysical Journal}. A portion is shown here for guidance
regarding its form and content.}
\tablenotetext{~}{References: (1) \citet{wolk96}; (2) \citet{zapatero02}; (3) \citet{barrado03}; (4) \citet{andrews04};
(5) \citet{kenyon05}; (6) \citet{caballero06}; (7) \citet{gonzalez08}; (8) \citet{sacco08}; (9) \citet{caballero12};
(10) \citet{alcala96}; (11) \citet{caballero06b}; (12) \citet{muzerolle03}; (13) \citep{maxted08};
(14) \citet{caballero08a}; (15) \citet{skinner08}; (16) \citet{lopez08}; (17) \citet{caballero09}; 
(18) \citep{caballero10}; (19) XMMN-SSC,2010); (20) \citet{burningham05}; (21) CVSO \citep{briceno05}; 
(22) Cluster Collaboration \citep{kenyon05,mayne07}; (23) \citet{cody10}; (24) AAVSO}
\tablenotetext{~}{Disk type: III=diskless, II=thick disk, I=class I, EV= evolved disk, TD=transition disk, DD=Debris disk}
\tablenotetext{~}{Member-flag: M: member, N: non-member, U: uncertain, P: probable member, U: uncertain member}
\tablenotetext{1}{\SOri AB:Multiple system}
\tablenotetext{2}{HD37699: probable disk detected using IRAS \citep{caballero07}}
\tablenotetext{3}{HD37333: It is a member if is a equal mass binary \citep{sherry08}}
\tablenotetext{4}{HD37564: It is too bright to be a cluster member \citep{sherry08}}
\tablenotetext{5}{HD294273: It is too faint to be a cluster member \citep[field star;][]{sherry08}}
\end{deluxetable}

\begin{deluxetable}{cccccccccccccc}
\rotate
\tabletypesize{\scriptsize}
\tablewidth{0pt}
\tablecaption{Membership for stars without spectral types studied in \S\ref{radialvelocity}\label{t:mem2}}
\tablehead{
\colhead{Name} & \colhead{2massID} & \colhead{Li} &  \colhead{Li} &\colhead{RV}  & \colhead{RV} & \colhead{H$\alpha$} & \colhead{disk} & \colhead{Xray} & \colhead{PM} & \colhead{Dist} & \colhead{Var} & \colhead{\%$_{pho}$}  & \colhead{Member}\\
\colhead{H07}    & \colhead{  } & \colhead{ flag } & \colhead{Ref}    & \colhead{flag}  & \colhead{Ref}  &  \colhead{flag}   &   \colhead{ } &  \colhead{Ref} &\colhead{flag}        & \colhead{flag} & \colhead{flag} & \colhead{ } & \colhead{flag} \\
}
\startdata
SO1154&05393982-0233159&0&4&\nodata&\nodata& Y&II&14,19 &2&1&21,22,23,24&22.7& M:\\
SO82&05373187-0245184& 0&\nodata&\nodata&\nodata& \nodata&III&\nodata&2&0&\nodata&22.9& U:\\
SO164&05374491-0229573& 0&\nodata&0&\nodata& \nodata&III& 14 &2&1&\nodata&10.6& N: \\
SO251&05375512-0227362& 2&\nodata&\nodata&\nodata& N&III&19&\nodata&1&\nodata&95.3& M:\\
SO302&05380167-0225527& 0&\nodata&\nodata&\nodata& N&III&16,18,19&\nodata&1&22&74.7& P:\\
SO304&05380221-0229556&0 &\nodata & 1 & \nodata&\nodata& III & \nodata & 1 & 1 &\nodata & 33.6 &  U: \\
SO329&05380561-0218571& 0&\nodata&0&\nodata& \nodata&III&\nodata&\nodata&1&\nodata&0.0& N:\\
SO371&05380966-0228569& 0&\nodata&\nodata&\nodata& \nodata&III&\nodata&\nodata&1&\nodata&0.0& N:\\
SO424&05381589-0234412& 0&\nodata&0      &\nodata& \nodata&III&\nodata&2      &1&\nodata&3.1& N:\\
SO482&05382307-0236493& 0&\nodata&\nodata& 20 & Y&II&\nodata&\nodata&1&21,22,23&50.5& M:\\
SO548&05382995-0215405& 0&\nodata&\nodata&\nodata& N&III&\nodata&\nodata&0&24&32.0& U:\\
SO561&05383138-0255032& 0&\nodata&0&\nodata& \nodata&III&\nodata&2&1&\nodata&23.3& N:\\
SO620&05383654-0233127& 0&\nodata&\nodata&\nodata& \nodata&III&\nodata&2&1&\nodata&82.3&U: \\
SO674&05384159-0230289& 2&8&2& 8 & N&II&\nodata&2&1&21,22&98.2& M:\\
SO692&05384375-0252427& 0&5&\nodata& 13,5& N&III&\nodata&\nodata&1&22&32.4& M:\\
SO738&05384809-0228536& 0&5&\nodata&5& N&II&\nodata&\nodata&1&\nodata&66.6& M:\\
SO773&05385173-0236033& 2&8&\nodata&8& N&III&14,15,18,19 &\nodata&1&21&31.1& M:\\
SO797&05385492-0228583& 0&5&\nodata& 5& N&III&14,19&\nodata&1&\nodata&55.7& M:\\
SO877&05390524-0233005& 0&5,6&\nodata& 5& N&III&14,18,19 &1&1&22,23&72.2& M:\\
SO917&05391001-0228116& 0&\nodata&\nodata&5&\nodata&EV&\nodata&\nodata&1&\nodata&3.6& M:\\
SO946&05391447-0228333& 0&3&2&8,2& N&III&14,18,19 &1&1&22,23&44.5& M:\\
SO1005&05392097-0230334&0&5&\nodata&5& N&III&14,19 &\nodata&1&21,22,23&4.9& M:\\
SO1043&05392561-0234042&2&\nodata&\nodata&\nodata& N&III&\nodata&\nodata&1&21,23&49.0& M:\\
SO1057&05392677-0242583&0&6&\nodata&\nodata&\nodata&EV&19&\nodata&1&22,23&8.1& M:\\
SO1370&05401304-0228314&0&\nodata&\nodata&\nodata& \nodata&III&\nodata&\nodata&0&\nodata&0.0& N:\\
\enddata
\tablenotetext{~}{References: (2) \citet{zapatero02}; (3) \citet{barrado03}; (4) \citet{andrews04};
(5) \citet{kenyon05}; (6) \citet{caballero06}; (8) \citet{sacco08}; (13) \citep{maxted08};
(14) \citet{caballero08a}; (15) \citet{skinner08}; (16) \citet{lopez08}; 
(18) \citep{caballero10}; (19) XMMN-SSC,2010); (20) \citet{burningham05}; (21) CVS; 
(22) Cluster Collaboration \citep{kenyon05,mayne07}; (23) \citet{cody10}; (24) AAVSO}
\tablenotetext{~}{Disk type: III=diskless, II=thick disk, I=class I, EV= evolved disk, TD=transition disk, DD=Debris disk}
\tablenotetext{~}{Member-flag: M: member, N: non-member, U: uncertain, P: probable member, U: uncertain member}
\end{deluxetable}

\begin{deluxetable}{cccccccccc}
\tabletypesize{\scriptsize}
\tablewidth{0pt}
\tablecaption{New photometric candidates with infrared excess at 24\micron \label{t:disk_mips}}
\tablehead{
\colhead{Name} & \colhead{2massID} & \colhead{V} & \colhead{[3.6]} & \colhead{[4.5]}  & \colhead{[5.8]} & \colhead{[8.0]} & \colhead{[24]} & \colhead{Disk} \\
\colhead{ }    & \colhead{  } & \colhead{mag} & \colhead{mag} & \colhead{mag} & \colhead{mag} & \colhead{mag} & \colhead{mag} & \colhead{type}  \\
}
\startdata
\nodata & 05373456-0255588 & 22.70 & \nodata        & 13.45$\pm$0.03 & \nodata        & 12.36$\pm$0.04 &  8.73$\pm$0.04 & II \\
SO238     & 05375398-0249545 & 18.47 & 11.11$\pm$0.03 & 10.72$\pm$0.03 & 10.39$\pm$0.03 &  9.79$\pm$0.03 & 7.24$\pm$0.03 & II  \\
\nodata & 05381279-0212266 & 20.28 & 12.93$\pm$0.03 &  \nodata       & 12.30$\pm$0.04 & \nodata        &  8.27$\pm$0.03 & II  \\
\nodata & 05382503-0213162 & 17.44 & 11.83$\pm$0.03 &  \nodata       & 11.34$\pm$0.03 & \nodata        &  8.39$\pm$0.03 & II  \\
\nodata & 05382656-0212174 & 15.12 & 10.18$\pm$0.03 &  \nodata       &  9.31$\pm$0.03 & \nodata        &  6.77$\pm$0.03 & II   \\
SO595     & 05383444-0228476 & 11.30 & 10.42$\pm$0.03 & 10.29$\pm$0.03 & 10.15$\pm$0.03 & 9.95$\pm$0.03  &  8.27$\pm$0.03 & DD/EV \\
\nodata & 05383981-0256462 & 14.62 & \nodata        &  9.31$\pm$0.03 & \nodata        & 7.90$\pm$0.03  &  5.05$\pm$0.03 & II  \\
\nodata & 05384714-0257557 & 21.15 & \nodata        & 12.82$\pm$0.03 & \nodata        & 11.55$\pm$0.03 &  8.94$\pm$0.04 & II    \\
\nodata & 05392639-0215034 & 14.51 & 9.33$\pm$0.03  & \nodata        & 8.38 $\pm$0.03 &  \nodata       &  4.18$\pm$0.03 & II    \\
SO1084    & 05393136-0252522 & 12.50 & 10.70$\pm$0.03 & 10.77$\pm$0.03 & 10.74$\pm$0.03 &  10.68$\pm$0.03 &  9.88$\pm$0.06 & DD/EV   \\
\nodata & 05394097-0216243 & 18.01 & 11.05$\pm$0.03 & \nodata        & 10.32$\pm$0.03 &  \nodata        &  7.57$\pm$0.03 & II  \\
\nodata & 05394278-0258539 & 14.25 &  \nodata       & 8.97$\pm$0.03  &  \nodata       & 7.75$\pm$0.03  &  4.43$\pm$0.03 & II    \\
SO1340    & 05400477-0245245 & 18.87 & 12.87$\pm$0.04  & 12.83$\pm$0.04 & 12.63$\pm$0.05 & 11.51$\pm$0.04 &  9.13$\pm$0.04 & II \tablenotemark{1}  \\
\nodata & 05400676-0257389 & 18.67 &  \nodata  &  \nodata &   \nodata & \nodata   &  9.58$\pm$0.05 & II\tablenotemark{1}  \\
\enddata
\tablenotetext{~}{Disk type: II=thick disk, DD/EV= debris disks or evolved disk}
\tablenotetext{1}{Potential galaxies based on the profile classification of UKIDSS}
\end{deluxetable}

\begin{deluxetable}{cccccccccccccc}
\rotate
\tabletypesize{\scriptsize}
\tablewidth{0pt}
\tablecaption{Known members compiled in \S\ref{known members} and  new disk bearing candidates not listed in Table \ref{t:mem1} and Table \ref{t:mem2} \label{t:appA}}
\tablehead{
\colhead{Name} & \colhead{2massID} & \colhead{V} &  \colhead{Disk} &\colhead{Li}  & \colhead{RV} & \colhead{Xray} & \colhead{PM} & \colhead{Dist} & \colhead{Var} & \colhead{\%$_{pho}$} & \colhead{Spectral} & \colhead{SpT}  & \colhead{Notes}\\
\colhead{H07}    & \colhead{  } & \colhead{ mag } & \colhead{Type}    & \colhead{Ref}  & \colhead{Ref}  &  \colhead{ref}   &   \colhead{flag} &  \colhead{flag} &\colhead{Ref} & \colhead{ } & \colhead{Type} & \colhead{ Ref} & \colhead{ } \\
}
\startdata
\nodata & 05401308-0230531 & 9.213 & \nodata & \nodata & \nodata & \nodata & 2       & 0 & \nodata & 17.1 & B9.5,B9.5 & 32,33 & \tablenotemark{2} \\ 
SO164 & 05374491-0229573 & 10.946 & III & \nodata & \nodata & 14 & 2 & 1 & \nodata & 10.6 & G9.0 & 14& \\
SO1084 & 05393136-0252522 & 12.500 & EV/DD & \nodata & \nodata & \nodata & 1 & 0 & \nodata & 0.0 & \nodata & \nodata& \tablenotemark{4}\\
\nodata & 05400365-0216461 & 14.946 & \nodata & 6 & \nodata & \nodata & 2 & 0 & 22 & 74.7 & \nodata & \nodata& \\
SO663 & 05384053-0233275 & 17.649 & II & 8 & 8,13 & \nodata & \nodata & 1 & 21 & 32.8 & M4.0 & 8& \\
SO1238 & 05395056-0234137 & 18.394 & III & 5 & 5,13 & \nodata & \nodata & 1 & 22,23 & 68.2 & M3.5 & 29& \\
\nodata & 05394299-0213333 & 18.580 & \nodata & 5 & \nodata & \nodata & \nodata & 0 & \nodata & 44.4 & M4.0 & 29& \\
SO976 & 05391699-0241171 & 18.862 & III & \nodata & \nodata & 14,18,19 & \nodata & 1 & 22,23 & 15.0 & \nodata & \nodata& \\
SO1005 & 05392097-0230334 & 19.255 & III & 5 & \nodata & 14,19 & \nodata & 1 & 21,22,23 & 4.9 & M5.0 & 29& \\
SO762 & 05385060-0242429 & 19.418 & II & 5 & 5,13 & \nodata & \nodata & 1 & 22 & 52.4 & M4,M5 & 29,30& \\
SO767 & 05385100-0249140 & 20.040 & III & 5 & 5 & \nodata & \nodata & 1 & \nodata & 6.9 & M4.5 & 29& \\
SO1019 & 05392319-0246557 & 21.060 & III & 5 & \nodata & \nodata & \nodata & 1 & \nodata & 25.3 & M5.5 & 9& \\
\nodata & 05374557-0229585 & \nodata & \nodata & 6 & \nodata & \nodata & 2 & 1 & \nodata & 0.0 & \nodata & \nodata&  \tablenotemark{3}\\
SO1215 & 05394741-0226162 & \nodata & III & \nodata & \nodata & 18 & 2 & 1 & \nodata & 0.0 & \nodata & \nodata&  \tablenotemark{3}\\
\enddata
\tablecomments{Table \ref{t:appA} is published in its entirety in the electronic edition of
the {\it Astrophysical Journal}. A portion is shown here for guidance
regarding its form and content.}
\tablenotetext{~}{References: (1) \citet{wolk96}; (2) \citet{zapatero02}; (3) \citet{barrado03}; (4) \citet{andrews04};
(5) \citet{kenyon05}; (6) \citet{caballero06}; (7) \citet{gonzalez08}; (8) \citet{sacco08}; (9) \citet{caballero12};
(12) \citet{muzerolle03}; (13) \citep{maxted08};
(14) \citet{caballero08a}; (15) \citet{skinner08}; (16) \citet{lopez08}; (17) \citet{caballero09}; 
(18) \citet{caballero10}; (19) XMMN-SSC,2010); (20) \citet{burningham05}; (21) CVSO \citep{briceno05}; 
(22) Cluster Collaboration \citep{kenyon05,mayne07}; (23) \citet{cody10}; (24) AAVSO;
(25) \citet{bejar99}; (26) \citep{scholz04}; (27) \citet{cody11};(28) \citet{rigliaco12}; 
(29) \citet{oliveira06}; (30) \citet{gatti08}; (31) \citet{oliveira04}; (32) \citet{sherry08};
(33) \citet{houk99} }
\tablenotetext{~}{Disk type: III=diskless, II=thick disk, I=class I, EV= evolved disk, TD=transition disk, DD=Debris disk}
\tablenotetext{1}{Source labeled as galaxy by \citet{lawrance13}}
\tablenotetext{2}{Bright member of the cluster \citep{sherry08,caballero07}}
\tablenotetext{3}{Photometric candidate \citep{lodieu09}}
\tablenotetext{4}{New disk bearing candidate (\S\ref{s:mips})}
\end{deluxetable}

\begin{deluxetable}{cccccccccc}
\tabletypesize{\scriptsize}
\tablewidth{0pt}
\tablecaption{Additional photometric candidates not listed in Table \ref{t:mem1}, Table \ref{t:mem2} and Table \ref{t:appA} \label{t:appB}}
\tablehead{
\colhead{Name} & \colhead{2massID} & \colhead{V} & \colhead{PM} & \colhead{Dist} & \colhead{Var} & \colhead{\%$_{pho}$} & \colhead{Spectral} & \colhead{SpT} \\
\colhead{H07}    & \colhead{  } & \colhead{ mag } &   \colhead{flag} &  \colhead{flag} &\colhead{Ref} & \colhead{ } & \colhead{Type} & \colhead{ Ref} \\
}
\startdata
\nodata & 05372067-0249330 & 10.338 & 0 & 0 & \nodata & 71.8 & G0 & 34 \\
\nodata & 05401245-0252576 & 10.494 & 1 & 0 & \nodata & 89.6 & F8 & 34 \\
SO168 & 05374536-0244124 & 10.659 & 0 & 1 & \nodata & 88.6 &  G0 & 34  \\
\nodata & 05375781-0226335 & 10.744 & 0 & 1 & \nodata & 33.0 & G0 & 34  \\
\nodata & 05371881-0231364 & 10.847 & 0 & 0 & \nodata & 56.8 & G0 & 34  \\
SO1224 & 05394891-0229110 & 17.34 & \nodata & 1 & 21,23 & 39.1 & \nodata & \nodata \tablenotemark{1}\\
\nodata & 05402081-0224003 & 24.77 & \nodata & 0 & \nodata & 50.9 & \nodata & \nodata \\
SO1163 & 05394057-0225468 & \nodata & 1 & 1 & \nodata & \nodata & F2 & 6,34 \\
\nodata & 05385382-0244588 & \nodata & \nodata & 1 & \nodata & \nodata & \nodata & \nodata \tablenotemark{2}\\
SO1141 & 05393816-0245524 & \nodata & \nodata & 1 & \nodata & \nodata & \nodata & \nodata \tablenotemark{2}\\
\enddata
\tablecomments{Table \ref{t:appB} is published in its entirety in the electronic edition of
the {\it Astrophysical Journal}. A portion is shown here for guidance
regarding its form and content.}
\tablenotetext{~}{References: (3) \citet{barrado03}; (6) \citet{caballero06};(21) CVSO \citep{briceno05}; (22) Cluster Collaboration \citep{kenyon05,mayne07}; 
(23) \citet{cody10}; (24) AAVSO; (29) \citet{oliveira06}; (34) \citet{nesterov95} }
\tablenotetext{1}{Source labeled as galaxy by \citet{lawrance13}}
\tablenotetext{2}{Photometric candidate \citep{lodieu09}}
\end{deluxetable}


\begin{figure}
\epsscale{0.8}
\plotone{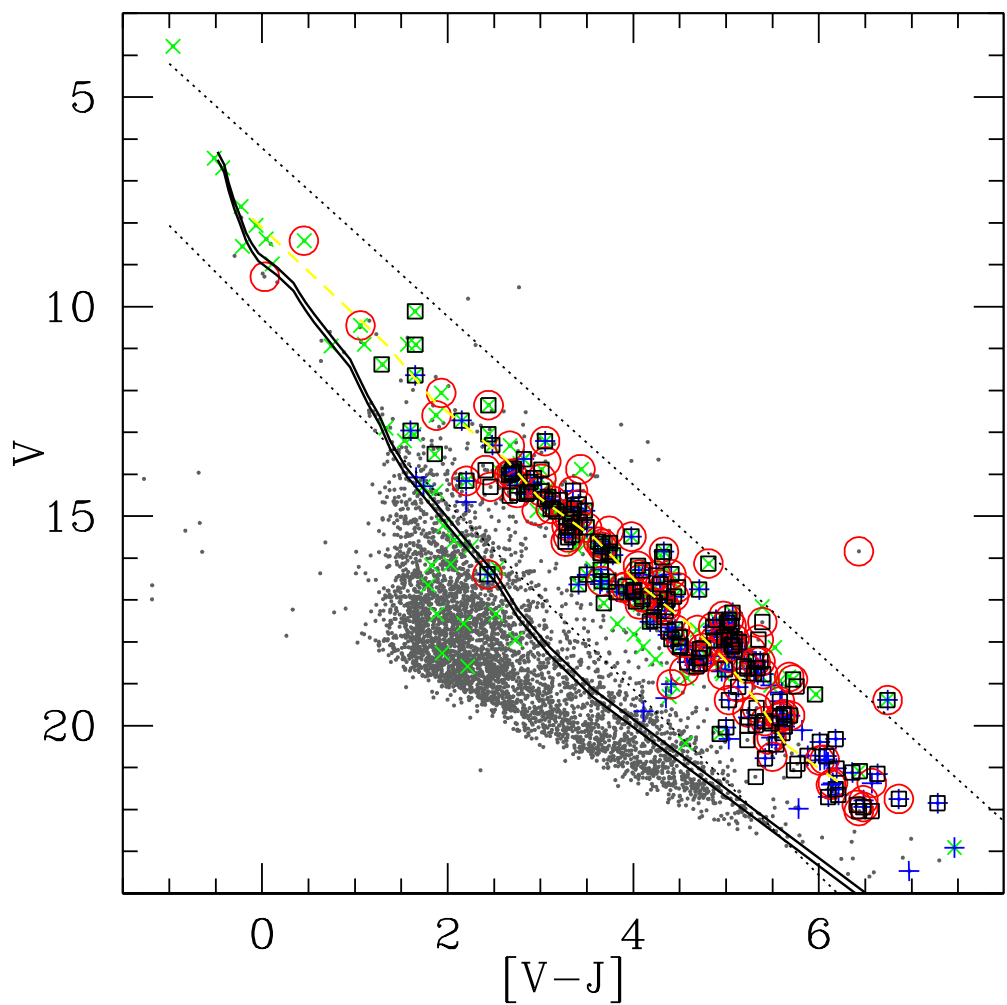}
\caption{ V versus V-J color magnitude diagram, illustrating the selection of photometric
members of the \SOri cluster. Open red circles represent stars with infrared excesses 
\citepalias{hernandez07a}.  Open  squares represent young stars confirmed using the 
presence of  \ion{Li}{1}$\lambda$6708 in absorption.
Plus signs represent kinematic members confirmed using radial velocities. 
Symbols "x" represent X ray sources. Likely members are X ray sources above the 
zero age main sequence (ZAMS).
The dashed line represents the empirical isochrone estimated from the median colors 
of the known members sample. Dotted lines limit the region where members are expected to fall 
(photometric member region). By comparison, the ZAMS from \citet{sf00} was plotted
assuming a distance of 420 pc \citep{sherry08} and 385 pc \citep{caballero08c}.
[See the electronic edition of the Journal for a color version of this figure.]}
\label{f:sel}
\end{figure}

\begin{figure}
\epsscale{0.8}
\plotone{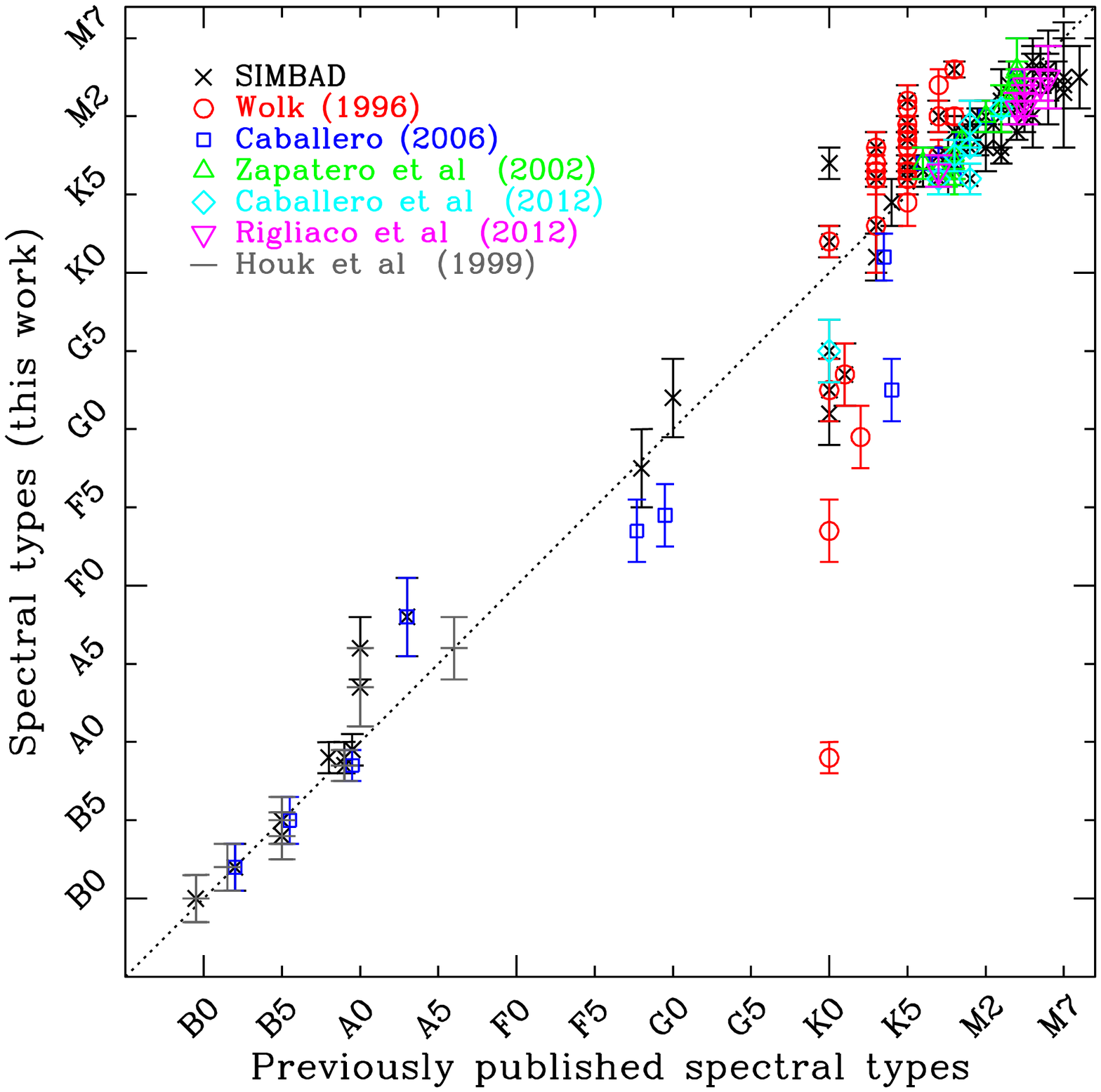}
\caption{Comparison of spectral types determined in this work with previously published values:
\citet{wolk96}, \citet{caballero06}, \citet{zapatero02}, \citet{caballero12}, \citet{rigliaco12}, \citet{houk99}
and SIMBAD database (see reference in the text).
Vertical error bars are the uncertainties derived from our spectral-type classification. 
For comparison, we show the line with slope 1.  For most stars, the spectral types derived in 
this work agree, within the uncertainties, with previous determinations of spectral types.
[See the electronic edition of the Journal for a color version of this figure.]}
\label{f:comp}
\end{figure}

\begin{figure}
\epsscale{0.8}
\plotone{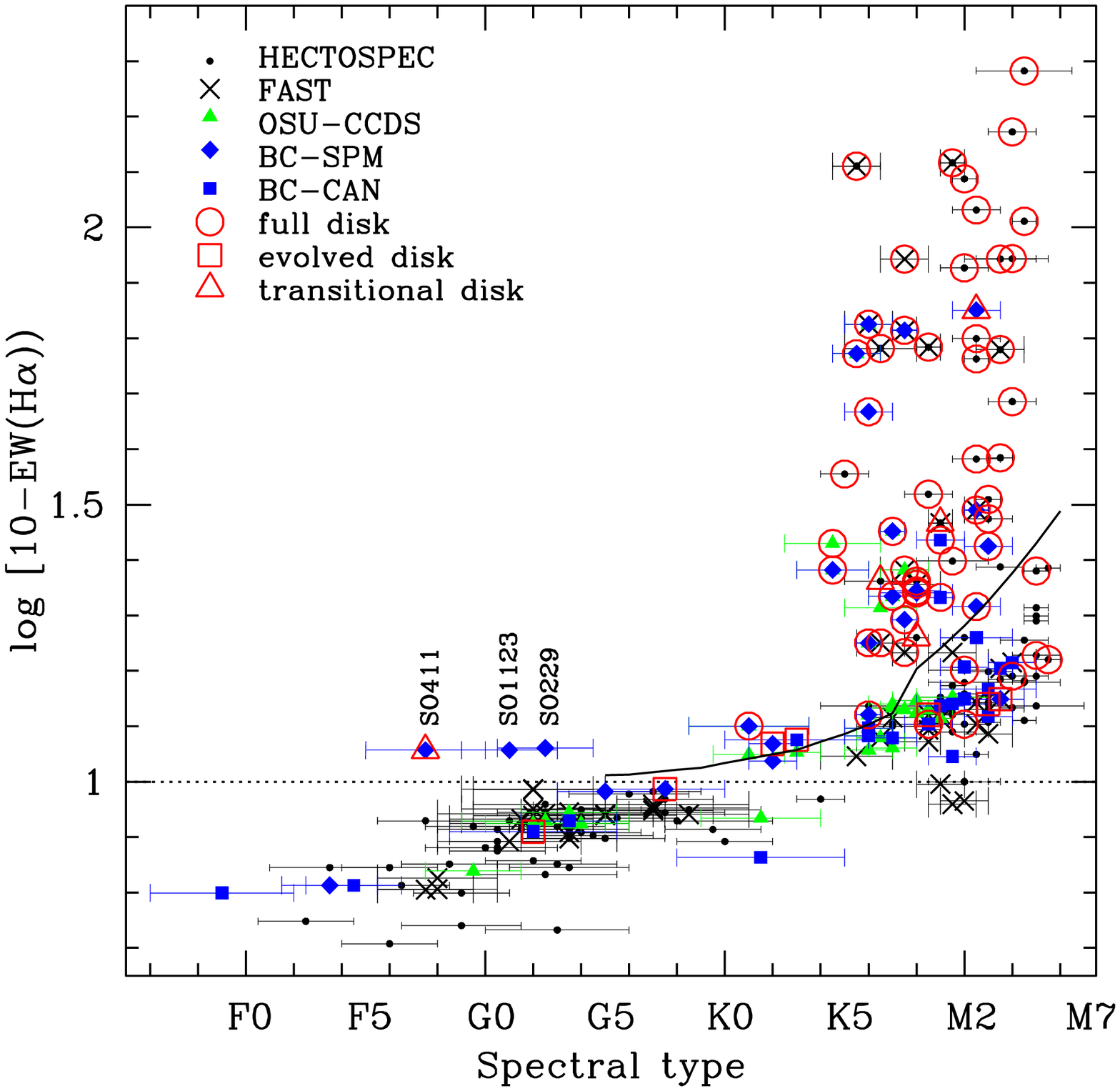}
\caption{Relation between the EW of H$\alpha$ and the spectral type. 
In order to plot our data in a logarithmic scale, EWs have been 
shifted by ten units. Solid line indicates the limit for Classical 
and Weak-line T Tauri stars based on the H$_\alpha$ in emission \citep{barrado03}.
Dotted line indicates the limit between emission and absorption of H$_\alpha$.
We plot with different symbols spectra obtained from different 
spectrographs (see Table \ref{t:spectrograph}). We also indicate 
the disk type for the sample using Spitzer photometry \citepalias{hernandez07a}.
[See the electronic edition of the Journal for a color version of this figure.]}
\label{f:CW_TTS}
\end{figure}

\begin{figure}
\epsscale{0.8}
\plotone{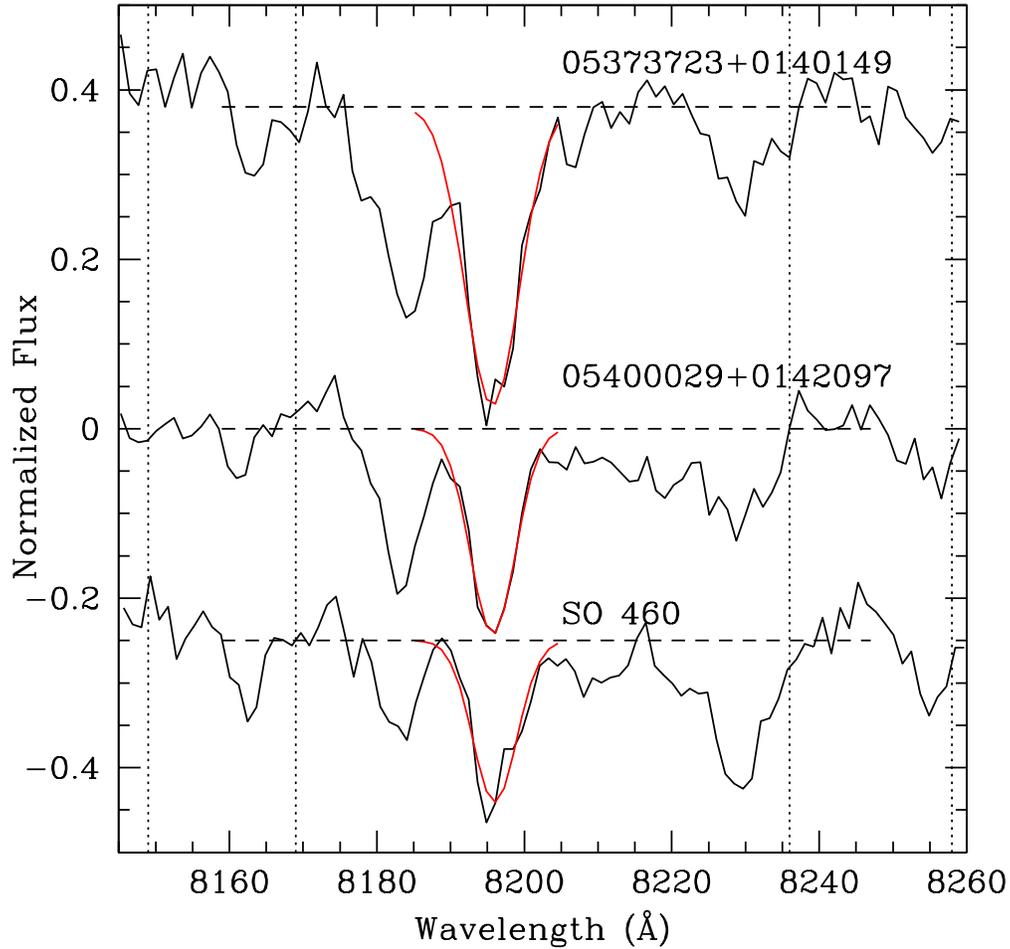}
\caption{Normalized spectra of the sodium doublet (\ion{Na}{1}$\lambda\lambda8183,8195$)
for three M5 type stars at different evolutionary stages. 
Vertical dotted lines represent the limits of the continuum bands used to normalize 
the spectra \citep{schlieder12}. A Gaussian function (blue line) was used to calculate 
the EW of the line \ion{Na}{1}$\lambda8195$. The M-type field dwarf (upper spectrum)
exhibits the strongest absorption of this feature. The PMS star (middle spectrum) located in the 
25 Ori stellar cluster \citep[age$\sim$10 Myr; ][]{briceno07} exhibits fainter \ion{Na}{1}$\lambda8195$
than the M type field dwarf, but stronger absorption than the stars in the \SOri cluster (lower spectrum). 
}
\label{f:sodium_lines}
\end{figure}

\begin{figure}
\epsscale{0.8}
\plotone{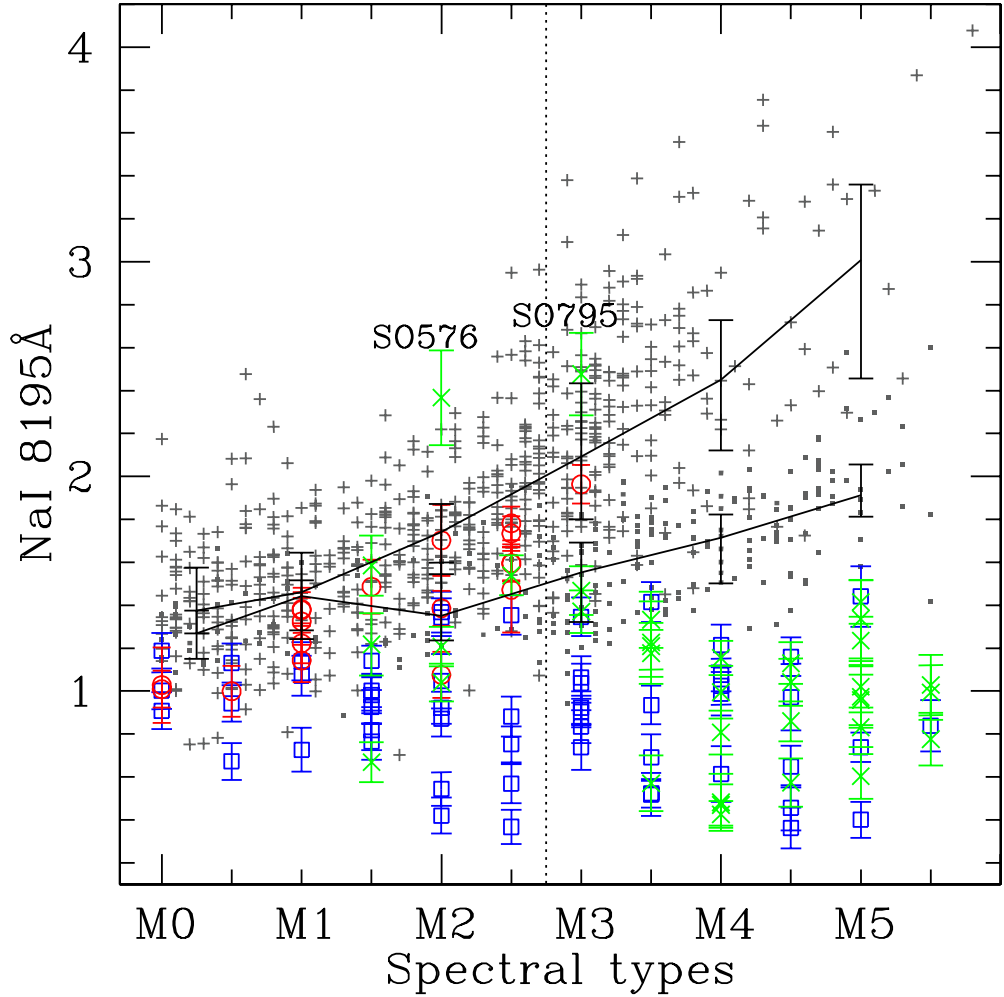}
\caption{Relation between the EW of \ion{Na}{1}$\lambda8195$ and the spectral type for the
sample of stars observed with Hectospec. We display the median and the second and third quartiles 
of \ion{Na}{1}$\lambda8195$ measured for a sample of PMS stars (age range $\sim$ 5-10 Myr; dots and 
lower solid line) and  for a sample of M-type field dwarfs (plus symbol and upper solid line) in the 
Orion OB1 association (Briceno in prep). Open squares and open circles indicate stars in the \SOri cluster with 
and without \ion{Li}{1}$\lambda$6708, 
respectively. In general, stars with uncertain membership based on \ion{Li}{1}$\lambda$6708 (crosses) have smaller values 
of EW of \ion{Na}{1}$\lambda8195$. Using the sodium criteria, we identify as member of the \SOri cluster
stars with spectral type M3 or later and located below the median values of the PMS population in 
the Orion OB1 association. The stars SO 576 and SO 795 were identified as M type field stars.
[See the electronic edition of the Journal for a color version of this figure.]}
\label{f:sodium_mem}
\end{figure}

\begin{figure}
\epsscale{0.8}
\plotone{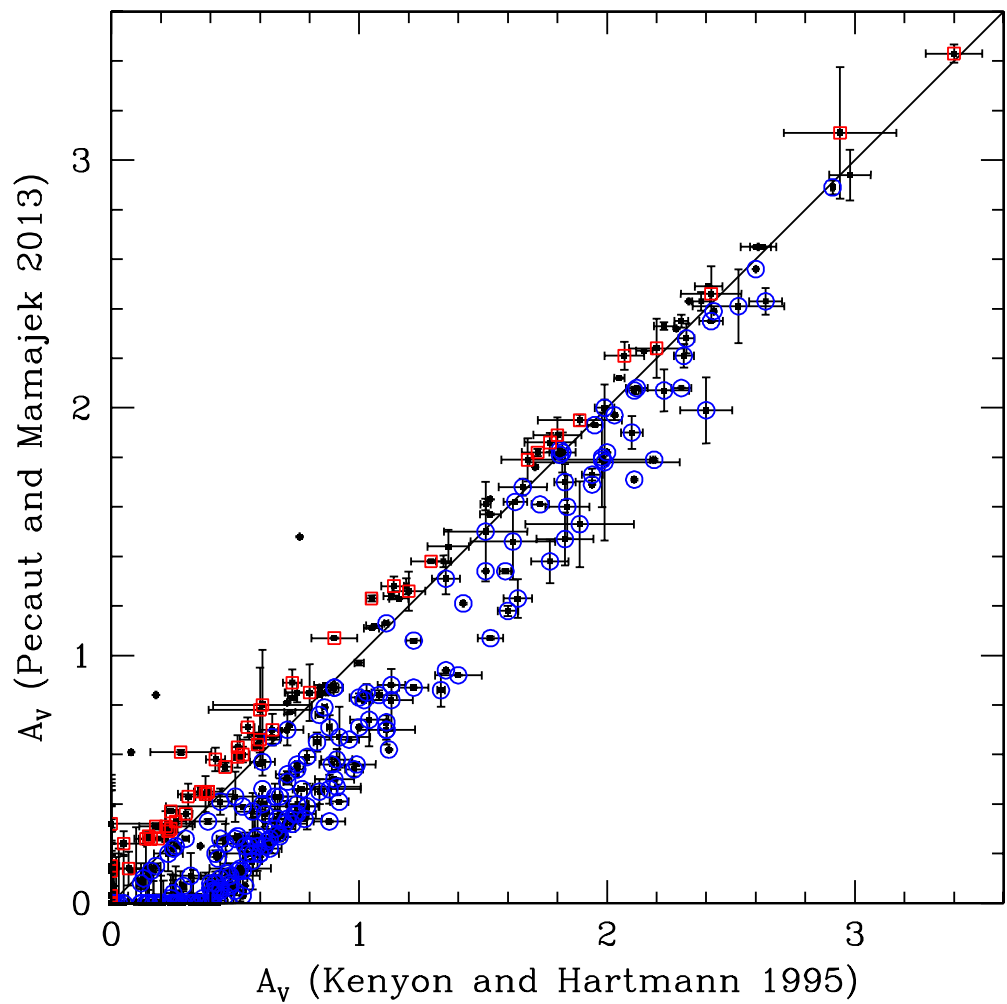}
\caption{Comparison between interstellar reddening calculated 
using the intrinsic colors for main sequence stars \citep{kh95}  
and the intrinsic colors for PMS stars \citep{pecaut13}. Open squares 
indicate stars with spectral type later than M3 and open circles indicate 
stars with spectral type range from G2 to M3. 
[See the electronic edition of the Journal for a color version of this figure.]}
\label{f:avcomp}
\end{figure}

\begin{figure}
\epsscale{0.8}
\plotone{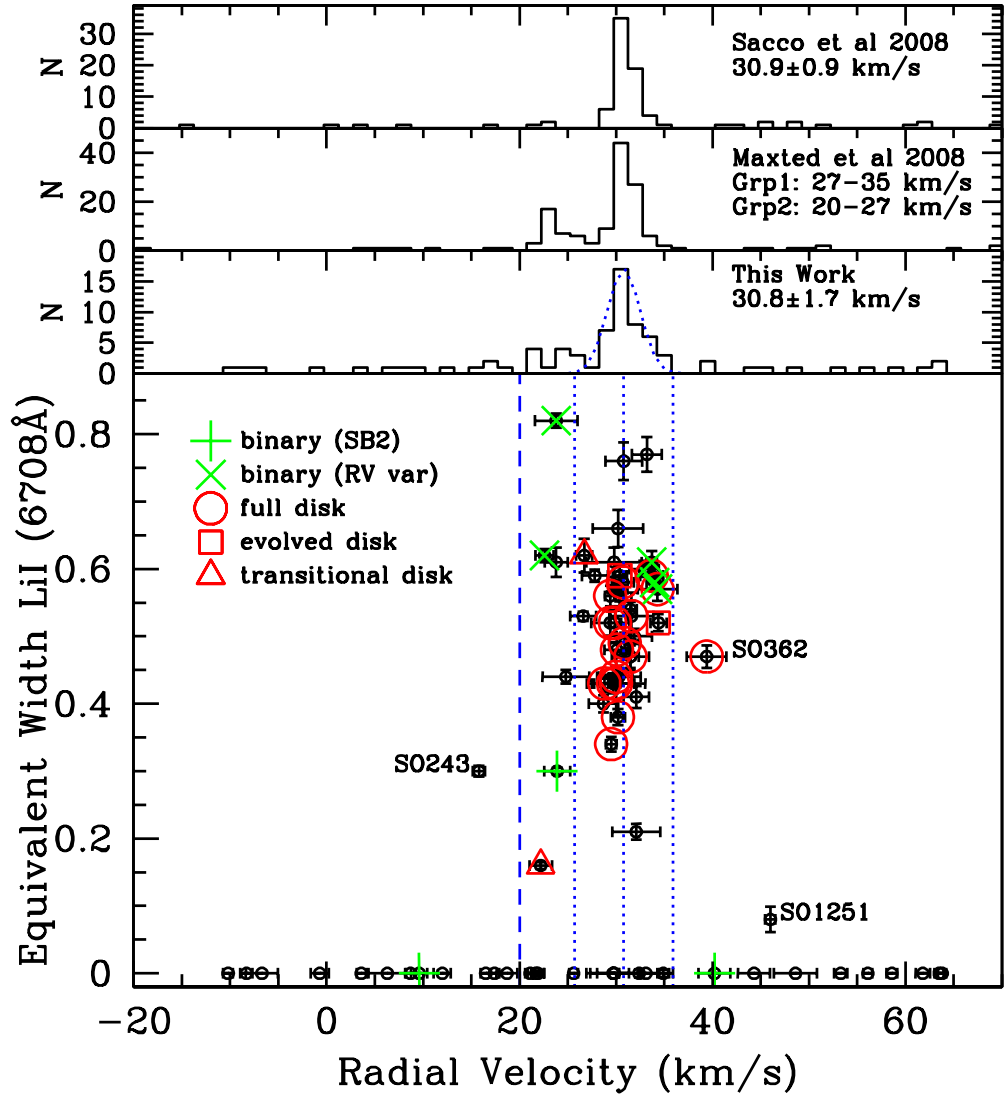}
\caption{Equivalent width of \ion{Li}{1}$\lambda$6708 versus Radial Velocity obtained from Hectochelle data.
Our RV distribution (third panel) is in agreement with the RV distribution obtained by \citet[][first panel]{sacco08} 
and with the RV distribution of the group 1 from \citet[][second panel]{maxted08}. A Gaussian function was fitted to 
our RV distribution. The dotted vertical lines represent the center of the Gaussian and the 3$\sigma$ criteria used
to identify  kinematic members of the cluster. The dashed line represents the lower limit for the sparser stellar population 
(see \ref{known members}). 
Disk types from \citetalias{hernandez07a} are plotted.  
We show binaries exhibiting double peaks in the correlation function (SB2) or binaries identified by radial velocity variability (RV var). 
Stars with \ion{Li}{1}$\lambda$6708 in absorption located to the left of the kinematic regions can be candidates 
of a more sparser and older group \citep[group 2 in][]{maxted08} or binary candidates of the cluster.
The stars SO 243, SO 362, SO 1251 are located in the kinematic non-members region.
[See the electronic edition of the Journal for a color version of this figure.]}
\label{f:RvLi}
\end{figure}

\begin{figure}
\epsscale{0.8}
\plotone{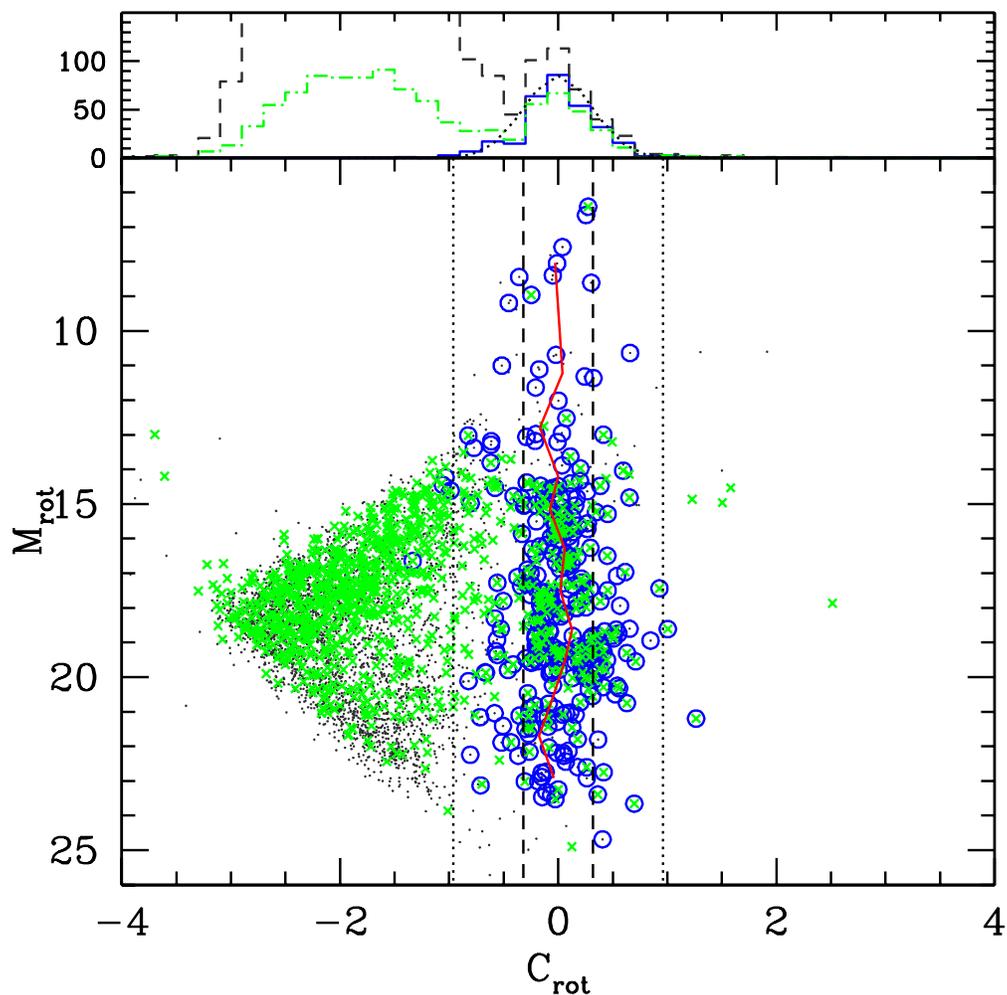}
\caption{Diagram used to calculate photometric probabilities. The color magnitude diagram of Figure \ref{f:sel}
was rotated in order to get the new variable C$_{rot}$ near zero for the known member sample 
(open circles and solid histogram in the upper panel). The red solid line represents the empirical isochrone. 
A Gaussian function (dotted histogram) was fitted to the C$_{rot}$ distribution of the known members.
Stars identified as variables 
are plotted with symbol "X". 
 Stars within the dotted vertical lines have photometric probabilities larger than 0.3\% (|C$_{rot}$|<3$\sigma$). Stars within 
the dashed vertical lines have photometric probabilities larger than 31.7\% (|C$_{rot}$|<1$\sigma$). 
The Dotted-dashed histogram and the
long dashed histogram represent the distribution 
of variable stars and the distribution of the entire sample, respectively.
[See the electronic edition of the Journal for a color version of this figure.]}
\label{f:CMD_rot}
\end{figure}

\begin{figure}
\epsscale{0.8}
\plotone{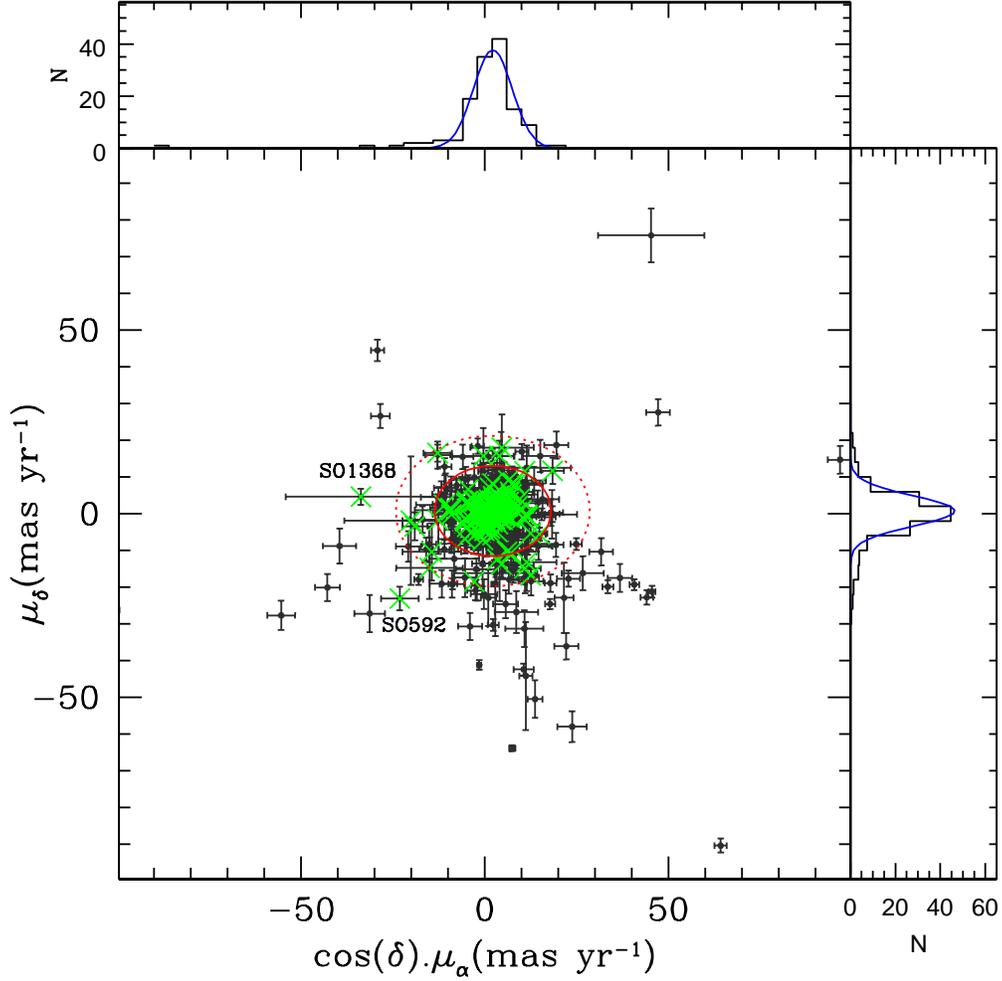}
\caption{Vector point diagram for photometric candidates. Green crosses indicate
the position of the known member sample, and the upper panel and right panel show
its proper motion distribution in right ascension and declination, respectively. 
Gaussian function curve fitting indicates that those distributions  are centered 
at  $cos(\delta)*\mu_\alpha \sim$ 2.2$\pm$5.3 mas/yr and $\mu_\delta \sim$0.7$\pm$4.1 mas/yr.
About 86\% of the known members are located within the 3$\sigma$ limit (dotted ellipse).
Almost all known members are located within 5$\sigma$ (solid ellipse) from the center 
of the distributions. Only the known members SO592, SO936 and SO1368 are located beyond
the 5 $\sigma$ criteria.
 [See the electronic edition of the Journal for a color version of this figure.]}
\label{f:PM}
\end{figure}

\begin{figure}
\epsscale{0.8}
\plotone{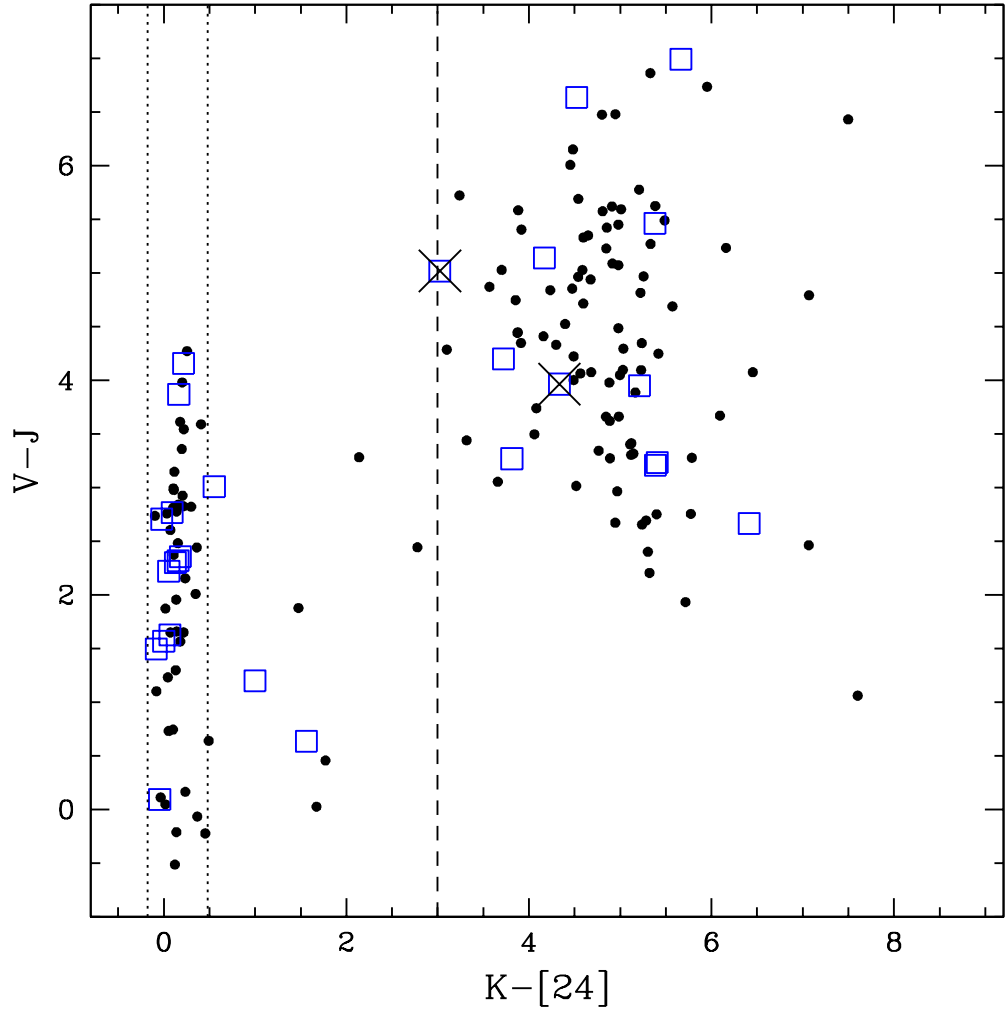}
\caption{Color magnitude diagram K-24 versus V-J 
illustrating the detection of disks at 24\micron. 
Circles and  squares represent sources included and not included 
in the previous disk census \citepalias{hernandez07a}, respectively. Dotted lines 
represent the photospheric K-24 colors estimated by \citetalias{hernandez07a}. 
The dashed line indicates an arbitrary limit between optically thick disks 
and debris disks estimated using a sample of debris disks located in  
several young stellar groups \citep{hernandez11}. Red X's are sources
labeled as galaxies by \citet{lawrance13}.
[See the electronic edition of the Journal for a color version of this figure.]}
\label{f:mips}
\end{figure}

\begin{figure}
\epsscale{0.8}
\plotone{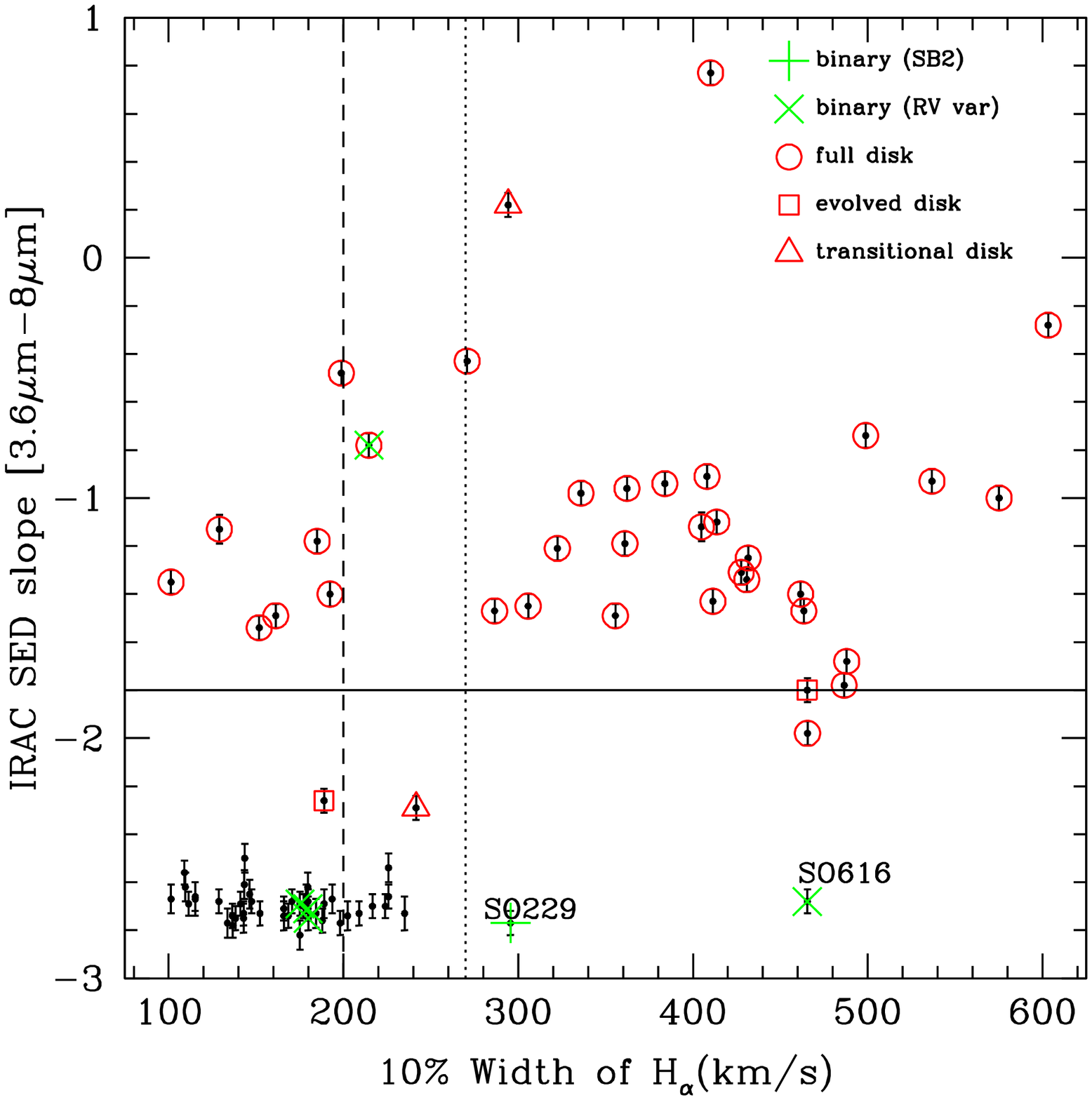}
\caption{IRAC SED slope versus the full width of H$_\alpha$ at 10\% of the line peak.
Dotted line and dashed line represent the limit between accretor and non accretor from
\citet{white03} and \citet{jayawardhana03}, respectively. Solid line indicates the 
limit between optically thick disks and evolved disks \citep{lada06}. Other symbols 
are similar to Figure \ref{f:RvLi}. The disk less stars SO 299 and SO 616 can be binaries or
fast rotators (see Figure \ref{f:rotator}). We identify 8 very slow accretors with optically thick disks. We also identify
one transitional disk (SO 818) and one evolved disk (SO 905) with accretion below the measurable
levels.
[See the electronic edition of the Journal for a color version of this figure.]}
\label{f:EW10alfa}
\end{figure}

\begin{figure}
\epsscale{0.8}
\plotone{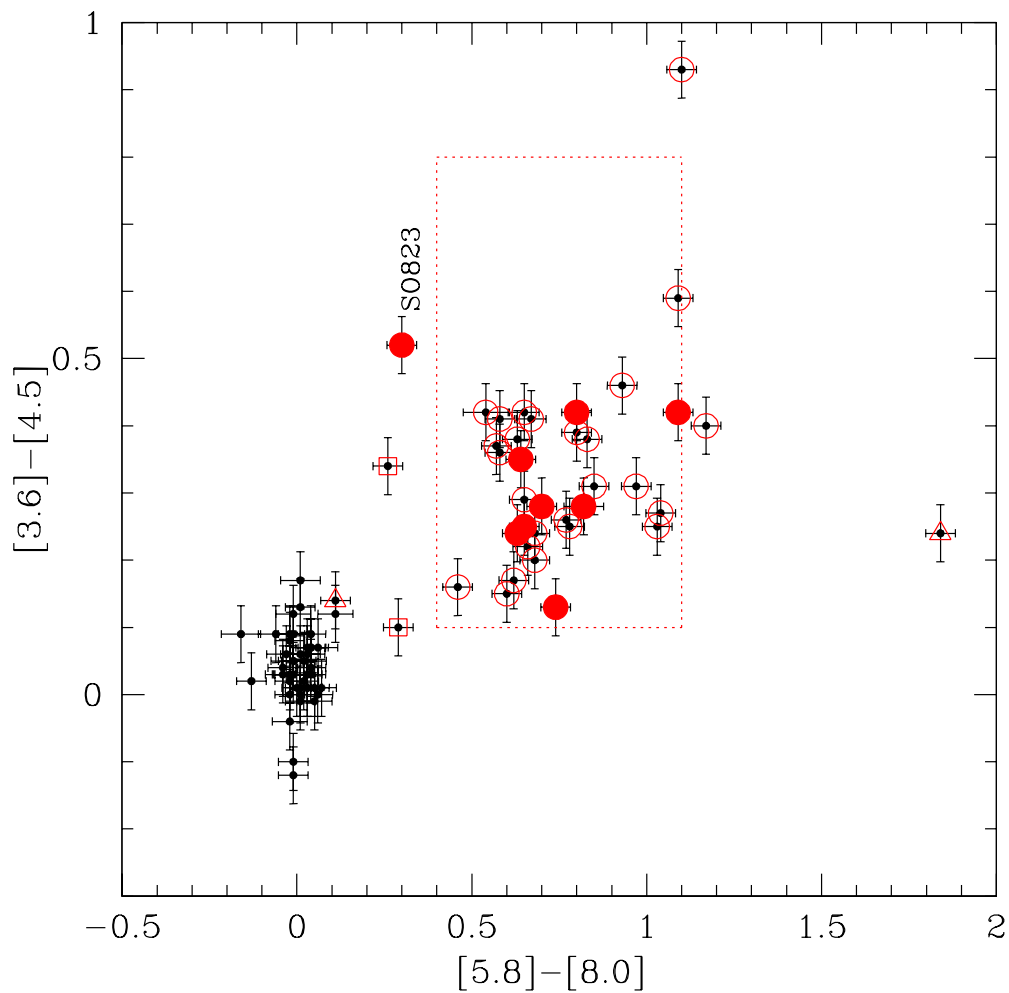}
\caption{IRAC color-color diagram [5.8]-[8.0] vs. [3.6]-[4.5] for stars included in Figure \ref{f:EW10alfa}.
Solid circles represent the very slow accretors candidates. We also included in this plot the 
slow accretors candidate SO 823. The large dotted box represents the loci of classical T Tauri stars 
(CTTSs) with different accretion rates \citep{dalessio06}. Very slow accretors candidates and 
stars with optically thick disks (Class II) which are above the accretion cutoff \citep{white03}  
share the same region in this plot. Symbols are similar to Figure \ref{f:EW10alfa}. 
[See the electronic edition of the Journal for a color version of this figure.]}
\label{f:irac}
\end{figure}

\begin{figure}
\epsscale{0.8}
\plotone{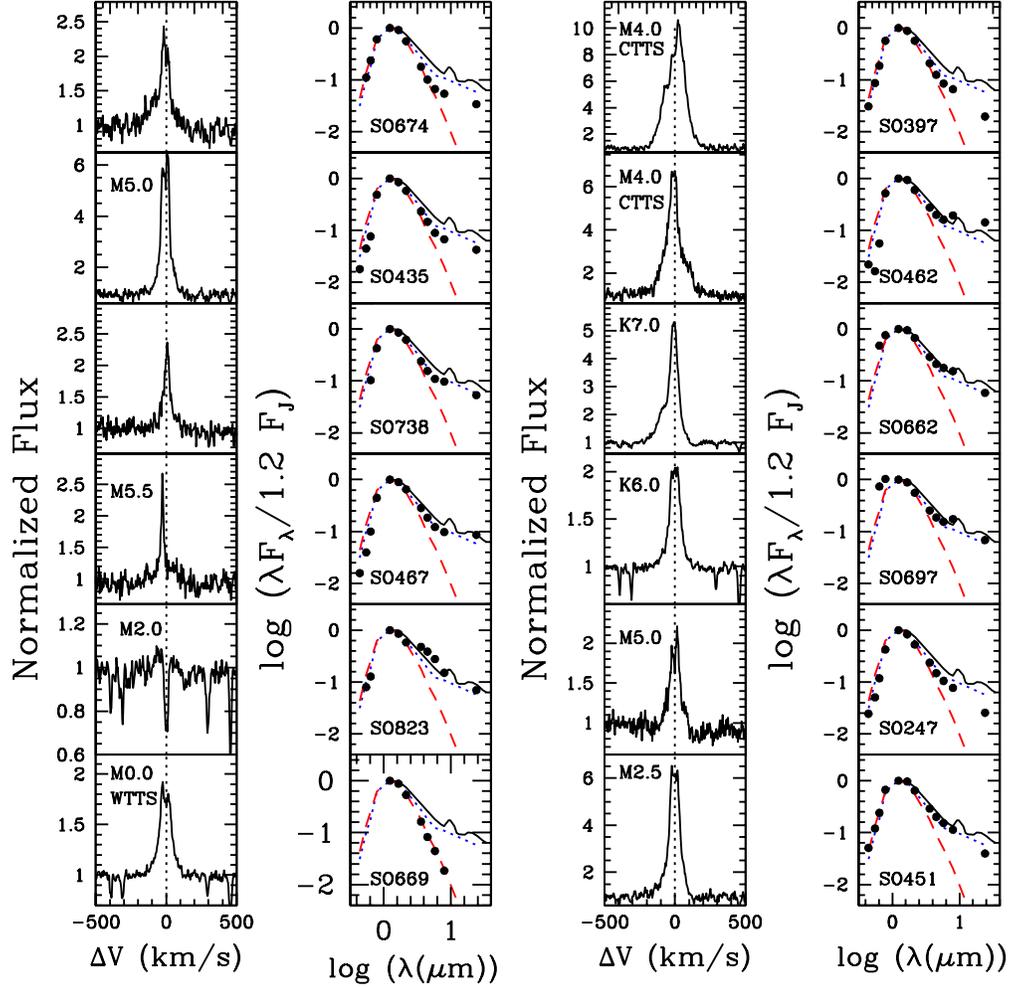}
\caption{Spectral energy distributions (SEDs) and H$_\alpha$ profiles for stars identified as 
very slow accretor candidates. For comparison we include a diskless star (SO 669) and two 
stars with accretion disks (SO 397 and SO 462). All SEDs are normalized at the J band. 
Values of the WH$_\alpha$\_10\% in km $s^{-1}$ are included in the panels that show the profile 
of H$_\alpha$. We do not measure the 
WH$_\alpha$\_10\% of the star SO 823 which exhibit a strong absorption component in the 
H$_\alpha$ line. In the panels that show the SEDs,  we plot the median SED for Class 
II stars in Taurus \citep[][solid lines]{furlan06}, the median SED for Class II (dotted lines) 
and the median SED for Class III (dashed lines) in the \SOri cluster \citepalias{hernandez07a}.
[See the electronic edition of the Journal for a color version of this figure.]}
\label{f:sed_Ha}
\end{figure}

\begin{figure}
\epsscale{0.8}
\plotone{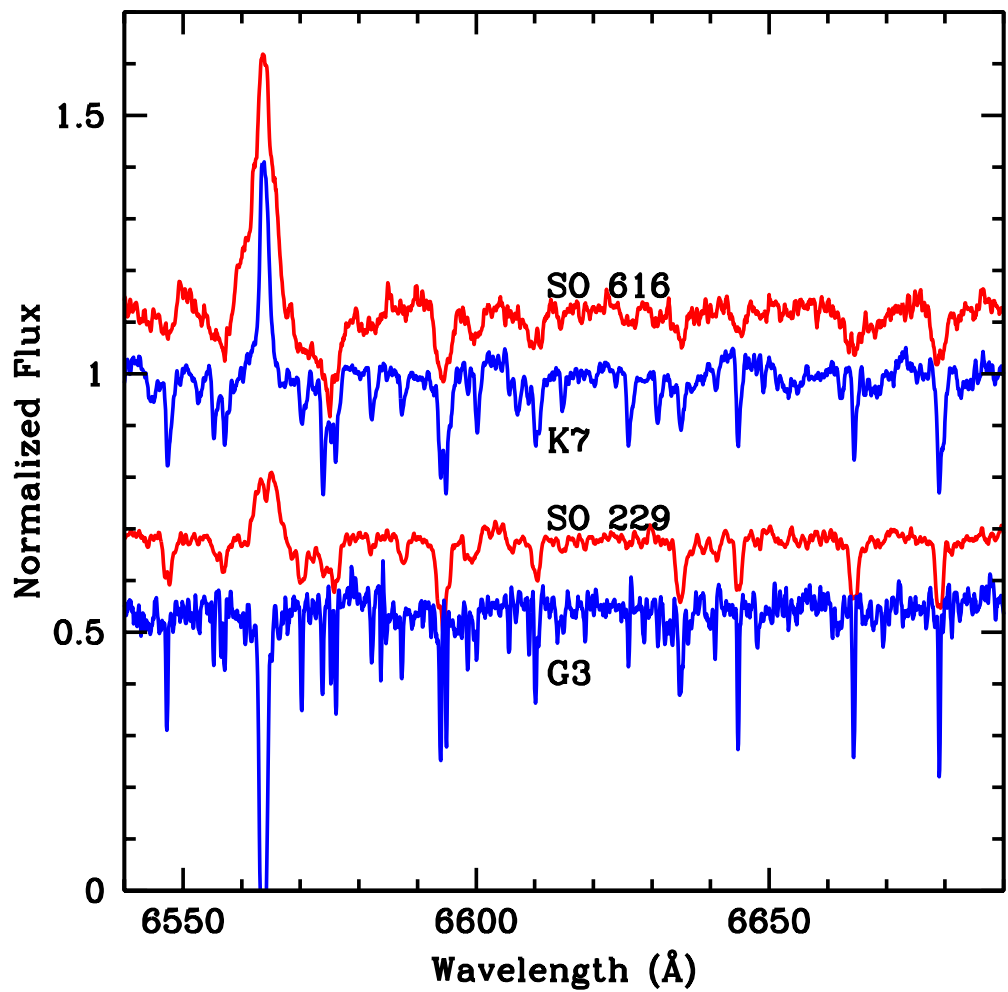}
\caption{Hectochelle spectra of the diskless stars SO 229 and SO 616 which mimic stars with accretion disks.
For comparison we show hectochelle spectra of  disk less stars with similar spectral types. Clearly, SO 229 and  SO 616 have
wider photospheric features in comparison to those stars. The broader spectral features could be produced by combined 
spectral lines of stellar components with similar effective temperature or by fast rotation.   
 }
\label{f:rotator}
\end{figure}

\end{document}